\documentclass[reqno]{amsart}
\input{preamble.sty}

\title[Higher tensor network states]{Higher tensor network states in two spatial dimensions: fundamental theorem for open boundary conditions}

\author{Clement Delcamp}
\address{Laboratoire Alexander Grothendieck, Institut des Hautes \'Etudes Scientifiques \& CNRS, Bures-sur-Yvette, France}

\author{Boris Kjær}
\address{Qmath, Department of Mathematical Sciences, University of Copenhagen, Denmark}

\thanks{\textit{E-mails}:
  \href{mailto:delcamp@ihes.fr}{\texttt{delcamp@ihes.fr}}, \href{mailto:bbk@math.ku.dk}{\texttt{bbk@math.ku.dk}}.}
 
\begin{document}

\begin{abstract}
    We introduce a class of two-dimensional tensor network states obtained by categorifying matrix product states. This procedure leads us to consider bond spaces provided by a 2-vector space associated with the 2-cells of the underlying lattice. Tensor network representations are then defined as module categories over the free monoidal category on a monoidal signature encoding the lattice geometry. We prove a fundamental theorem stating that two representations generating the same states with suitable open boundary conditions are equivalent as module categories. Finally, we introduce a notion of 2-injectivity, a categorified analogue of the injectivity condition for matrix product states.
\end{abstract}

\maketitle
\tableofcontents

\section{Introduction}

\noindent
On the face of it, solving quantum spin systems is hard, even in low spatial dimensions. But physical states are not random, and only cover a small corner of the exponentially large Hilbert space. Specifically, ground states of local \emph{gapped} Hamiltonians in one spatial dimension possess an entanglement entropy which obeys an \emph{area law} \cite{Hastings:2007iok}. Such states can be efficiently approximated by \emph{matrix product states} (MPS), a class of wave functions that implement the area law at a microscopic level, thereby allowing one to target the physical corner directly \cite{ACCARDI1981169,PhysRevLett.59.799,Affleck1988,Fannes1992,PhysRevB.73.094423}. The \emph{density matrix renormalisation group} (DMRG) algorithm, a variational algorithm within the space of MPS, takes full advantage of this specific entanglement pattern in order to overcome the exponential cost \cite{PhysRevLett.69.2863,SCHOLLWOCK201196,PhysRevLett.93.227205}. 

A \emph{uniform} MPS is specified by a single tensor, copies of which are contracted along so-called \emph{bond spaces} to produce the state. Interestingly, distinct tensors may represent the same state. For instance, this occurs when two tensors are related by a \emph{gauge transformation}, namely an invertible matrix acting on the bond space, which  cancels pairwise as the tensors are contracted. \emph{Fundamental theorems} address the converse question \cite{Perez-Garcia:2006nqo,PhysRevLett.100.167202,Cuevas:2017jlr,Florido-Llinas:2025hbb}: given two MPS representations of the same state, how must they be related? The variety of such theorems stems from the precise nature of the setup, the properties assumed of the tensors, and the strength of the hypothesis. 
The simplest theorems assume the defining tensor to be \emph{injective}, i.e., that the tensor, interpreted as a map from bond spaces to the physical local vector space, is injective. Mathematically, this case is particularly tractable; physically, the resulting states are unique ground states of local, frustration-free, gapped, parent Hamiltonians \cite{Fannes1992}. 
The corresponding fundamental theorems produce a classification of symmetric injective MPS that reproduces that of symmetry-protected topological (SPT) phases, both for invertible \cite{PhysRevLett.100.167202,Schuch:2011niz,PhysRevB.81.064439,PhysRevB.83.035107} and non-invertible symmetries \cite{Garre-Rubio:2022uum,Inamura:2021wuo}. 
More generally, provided that the tensors are brought into a so-called \emph{canonical form}, a general fundamental theorem can also be proven.

\bigskip \noindent
In two spatial dimensions, the situation is richer. Various families of tensor network states can be constructed, the most straightforward generalisation of MPS being so-called \emph{projected entangled pair states} (PEPS). These play an analogous role, albeit with weaker guarantees: numerical simulations are less efficient due to exact contractions being computationally hard, and a general fundamental theorem has been argued to be out of reach \cite{Scarpa:2018cbk}. Nevertheless, as in one spatial dimension, one can define injective PEPS, which were shown to be unique ground states of local parent Hamiltonians, the latter being either gapped or gapless \cite{Perez-Garcia:2007vgo}. A fundamental theorem has been proven in this case \cite{Perez-Garcia:2010maz}, as well as for \emph{normal} PEPS \cite{Molnar:2018zys}, which only become injective after blocking sufficiently many tensors together. A more general theorem has also been established for PEPS in a \emph{minimal canonical form} \cite{Acuaviva:2022lnc}. In all of these cases, the corresponding gauge transformations factorise as tensor products. 
Consequently, the classification of symmetric injective/normal PEPS fails to reproduce that of SPT phases for (ordinary) 0-form invertible symmetries.  This has led researchers to introduce new classes of tensor network states (TNS) and to impose new sets of conditions.   

On the one hand, there is a notion of \emph{semi-injective} TNS.\footnote{In contrast to the original paper, we refrain from referring to them as PEPS, since they possess a different type of \emph{entanglement structure} \cite{Christandl:2020chy}, and rather use the more general term `tensor network state'.} Rather than a property of the individual tensors, semi-injectivity refers to a particular structure of the tensor network, tied to a specific geometry. A fundamental theorem for semi-injective TNS was proven in ref.~\cite{Molnar:2018hls}, which yields a classification of symmetric semi-injective TNS matching that of SPT phases for 0-form invertible symmetries. On the other hand, there is a notion of \emph{MPO-injective} PEPS \cite{Buerschaper:2013yfm,Sahinoglu:2014upb,Williamson:2014ydz,Lootens:2020mso}, introduced as a generalisation of $G$-injectivity \cite{Schuch:2010hfh}, 
whereby the tensor is injective on a subspace of the bond space defined by a matrix product operator (MPO) projector. 
Although no general fundamental theorem is known for this more general class of PEPS, the corresponding MPO symmetries have also been shown to recover the classification of SPT phases for 0-form invertible symmetries. 
More recently, the same MPOs have been used to encode symmetry and duality operators in the context of one-dimensional quantum lattice models \cite{Lootens:2021tet,Lootens:2022avn,Lootens:2023wnl}. 

So far, no fundamental theorem for any  class of TNS has been shown to reproduce the broader classification of SPT phases in two spatial dimensions for \emph{higher group} symmetries, non-invertible symmetries, and combinations thereof, as would be encoded into a \emph{fusion 2-category}. Such a classification is expected to match that of \emph{fibre 2-functors} established in ref.~\cite{Decoppet:2023bay}, obtained by lifting results of Ostrik \cite{2002math......2130O} to the higher setting. Although, we do not address this specific question in the present work,\footnote{This will be the main result of a separate manuscript.} it does motivate our approach. 

\bigskip \noindent
In this work, we introduce a new class of tensor network representations, and prove a fundamental theorem for pairs of such representations generating the same states for a given choice of \emph{open} boundary conditions. For a specific choice of the underlying data, one recovers ordinary PEPS, while different choices yield TNS encoding distinct entanglement structures, in a way that is reminiscent of semi-injective TNS. 
Conceptually, we approach the definition of such tensor network representations as a \emph{categorification} problem \cite{baez1998categorification}. Categorification is a loosely defined procedure which amounts to replacing set-theoretic statements by category-theoretic analogues. 
In physics, it often accompanies an increase of the spatial dimension by one; going from one to two spatial dimensions, it typically amounts to replacing complex vector spaces by particularly well-behaved categories referred to as \emph{2-vector spaces} \cite{kapranov19942,BAEZ1997125}.\footnote{Viewing complex vector spaces as modules over the ring $\mathbb C$, a 2-vector space may be defined as a module category over the category $\Vect$ of finite-dimensional complex vector spaces. In this manuscript, we further require such a module category to be finite and semisimple.} Structures on vector spaces are then promoted accordingly, albeit not in a unique way; for instance, the notion module over an algebra may be promoted to that of a \emph{module category} over a \emph{monoidal category}.

Applying the above philosophy to tensor networks first requires reformulating, in a more abstract way, what a (uniform) tensor network representation in one spatial dimension is. In this manuscript, we define it as a finite-dimensional module over the free algebra $F^d$ on the vector space $\mathbb C^{\oplus d}= \Span_\mathbb C\{e_i \, | \, i\in [d]\}$, for some positive integer $d$. Here, the free algebra $F^d$ is used to encode Hilbert spaces of the form $(\mathbb C^{\oplus d})^{\otimes L}$, for arbitrary $L$. By freeness, such a module $V^A$ amounts to a $\mathbb C$-linear map $\mathbb C^{\oplus d} \to \End_\mathbb C(V^A), e_i \mapsto A^i$, which is nothing but the defining tensor. In this formulation, the fundamental theorem states that, under suitable hypotheses, two tensor network representations generating the same states are isomorphic as $F^d$-modules. A proof of such a theorem for closed periodic boundary conditions can already be found, in disguise, in ref.~\cite{MR1610851}. In this manuscript, we provide a proof for open boundary conditions, under an assumption that does not entail any loss of generality. This proof serves as a template for our two-dimensional treatment.  

\bigskip \noindent
We summarise below our main findings.
Categorifying the above treatment of the  one-dimensional case, we propose to replace the free algebra $F^d$ by a free (strict $\mathbb C$-linear) monoidal category on a \emph{monoidal signature} \cite{JOYAL199155,Selinger2011}.  
The monoidal signature encodes, in particular, a choice of lattice geometry. 
For a square geometry, one considers the 
free monoidal category $\Fsq$ generated by two objects $X_1,X_2$ and a $d$-dimensional space of morphisms $X_1X_2 \to X_2X_1$. 
Concretely, non-zero hom-spaces in $\Fsq$ are of the form $(\mathbb C^{\oplus d})^{\otimes N}$, where $N$ is the number of vertices of a specific kind of patch of the square lattice, connected such patches being referred to as \emph{shapes}. 
Similarly, we propose to replace the bond space by a categorified analogue, which we choose to  model as a 2-vector space.
\begin{introdefinition}
    We define a uniform tensor network representation on the square geometry with local Hilbert space $\mathbb C^{\oplus d}$ as a strict module category over the free strict $\mathbb C$-linear monoidal category $\Fsq$.
\end{introdefinition}

\noindent
A strict module category over $\Fsq$ consists of a 2-vector space $\mc M^A$ together with $\mathbb C$-linear strict monoidal functor $\rho^A \colon \Fsq \to \EndC(\mc M^A)$. By freeness, $\rho^A$ amounts to two endofunctors $T^A_1 := \rho^A(X_1)$ and $T^A_2 := \rho^A(X_2)$, together with a $\mathbb C$-linear map $\mathbb C^{\oplus d} \to \Nat(T^A_1 \circ T^A_2 , T^A_2 \circ T^A_1)$.  Unpacking this definition, a tensor network representation on the square geometry boils down to the following data: a set $\Irr(\mc M^A)$ of labels for the regions between bonds; bond spaces 
\begin{equation*}
    \VA1 \equiv  \!\! \bigoplus_{M_1,M_2 \in \Irr(\mc M^A)} \!\! (T^A_1)^{M_2}_{M_1} \q \text{and} \q \VA2 \equiv  \!\! \bigoplus_{M_2,M_4 \in \Irr(\mc M^A)} \!\! (T_2^A)_{M_2}^{M_4} \, ,
\end{equation*}
whose blocks are labelled by the two regions on either side of the bond; and a collection of $\mathbb C$-linear maps
\begin{align*}
    A^i 
    &\equiv \PEPS{1}{A}{i}{}{}{}{}{}{}  = \bigoplus_{M_1,\ldots,M_4} \,  \PEPS{1}{A}{i}{M1234}{}{}{}{}{} 
    \\
    &\equiv  \PEPSord{i} \in (V_2^A)^* \otimes (V_1^A)^* \otimes V_2^A \otimes V_1^A\, .
\end{align*}
The maps $A^i$ vanish outside the blocks with matching region labels. Labelling the regions by $M_1,\ldots, M_4 \in \Irr(\mc M^A)$ amounts to selecting the block $(T^A_1)^{M_4}_{M_2} \otimes (T_2^A)_{M_1}^{M_2} \to (T_2^A)_{M_3}^{M_4} \otimes (T_1^A)_{M_1}^{M_3}$. An $\Fsq$-module category $\mc M^A$ is said to be \emph{non-degenerate} whenever the $\Vect$-valued matrices with entries $(T_1^A)_{M_1}^{M_2}$ and $(T^A_2)_{M_2}^{M_4}$, respectively, have no vanishing rows or columns.  
 
We are interested in tensor network states on shapes with a special class of open boundary conditions specified by \emph{boundary data}, namely vectors $v_1^A \in V^A_1$, $v_2^A \in V^A_2$ and covectors $\phi_1^A \in (V_1^A)^*$, $\phi_2^A \in (V_2^A)^*$. 
Given a shape, we place a copy of $\vA1$ (resp. $\vA2$) at the end of every incoming vertical (resp. horizontal) leg, as well as a copy of $\phA1$ (resp. $\phA2$) at the end of every outgoing vertical (resp. horizontal) leg, and contract consecutive copies along the label of the region they share. This produces a boundary MPS whose bond carries region labels and is depicted below as a green line. The tensor network state is obtained by contracting copies of $A^i$, one for each vertex, along the edges of the shape, and closing off the remaining legs with the boundary MPS.

We say that the boundary vectors $(v_1^A,v_2^A)$ are cyclic if
\begin{equation*}
    \text{Span}_\mathbb C \bigg\{ \PEPSmps{i} \,\bigg|\; i \in [d] \, \bigg\} = \bigoplus_{M_1,M_3,M_4} (T^A_2)^{M_4}_{M_3} \otimes (T^A_1)^{M_3}_{M_1} =: V_{21}^A \, ,
\end{equation*}
where the incoming legs are closed off with the boundary MPS.
Dually, the boundary covectors $(\phi^A_1,\phi^A_2)$ are cocyclic if
\begin{equation*}
    \text{Span}_\mathbb C \bigg\{ \PEPSmpsco{i} \,\bigg|\; i \in [d] \, \bigg\} = \bigoplus_{M_1,M_2,M_4} \big((T^A_1)^{M_4}_{M_2} \otimes (T^A_2)^{M_2}_{M_1}\big)^* =: (V^A_{12})^* \, .
\end{equation*}
Finally, the boundary data $(v^A_1,v^A_2,\phi^A_1,\phi^A_2)$ is said to have \emph{full support} if these have non-zero components in every non-zero block of $\VA1$, $\VA2$, ${\VA1}^*$ and ${\VA2}^*$, respectively.
Our main  result in the following theorem:
    
\begin{introtheorem}
    Let $\mc M^A$ and $\mc M^B$ be  two non-degenerate $\Fsq$-module categories such that $\dim_\mathbb C V_{12}^B \leq \dim_\mathbb C V^A_{12}$, $\dim_\mathbb C V_{21}^B \leq \dim_\mathbb C V^A_{21}$. Moreover, let $(\vA1,\vA2,\phA1,\phA2)$ and $(\vB1,\vB2,\phB1,\phB2)$ be boundary data with full support such that $(\vA1,\vA2)$ is cyclic and $(\phA1,\phA2)$ is cocyclic. If, for every shape, $\mc M^A$ and $\mc M^B$ generate the same tensor network state with their respective boundary data, then there is an equivalence $(\mc Q,\zeta) \colon \mc M^A \to \mc M^B$ of $\Fsq$-module categories, unique up to isomorphism, that preserves the boundary data.
\end{introtheorem}
\noindent
Concretely, such an equivalence amounts to invertible maps $Q_1 \colon \VA1 \to \VB1$ and $Q_2 \colon \VA2 \to \VB2$ that relabel the regions according to a bijection $\Irr(\mc M^A) \cong \Irr(\mc M^B)$. These can be contracted along the labels of the regions they share so as to produce an MPO whose bond carries region labels and is depicted below as a double line. This MPO intertwines the two tensor network representations:
\begin{equation*}
    \PEPSQord{21}{i} = \PEPSQord{12}{i} \, ,
\end{equation*}
for every $i \in [d]$. Preserving the boundary data then means that $Q_1 \circ \vA1 = \vB1$, $Q_2 \circ \vA2 = \vB2$, $\phB1 \circ Q_1  = \phA1$ and $\phB2 \circ Q_2  = \phA2$. Although our focus in this manuscript is on the square geometry, our framework applies to any monoidal signature, and we expect analogous fundamental theorems to hold.

\bigskip \noindent
In order to make contact with the literature---and with a view towards future physical applications to (2+1)d SPT phases---let us mention a stronger assumption, which we regard as a categorification of the ordinary injectivity condition. We dub this condition \emph{2-injectivity}. Physically, this condition is meant to encode states that are unique ground states of local Hamiltonians. Mathematically, it ensures that any non-zero boundary data with full support is cyclic and cocyclic.

\begin{introdefinition}[2-injectivity] \label{def:2injectivity}
    A tensor network representation $\mc M$ on the square geometry is 2-injective if $(T_1)^{M_2}_{M_1}$ and $(T_2)^{M_2}_{M_1}$ are non-zero for every $M_1,M_2 \in \Irr(\mc M)$, and if 
    \begin{equation*}
        \Span_\mathbb C \{A^i \, | \, i \in [d]\} = \bigoplus_{M_1,\ldots,M_4} {(T_2)^{M_2}_{M_1}}^* \otimes {(T_1)^{M_4}_{M_2}}^* \otimes (T_2)^{M_4}_{M_3} \otimes (T_1)^{M_3}_{M_1} \, .
    \end{equation*}
\end{introdefinition}

\noindent
First of all, note that whenever the 2-vector space $\mc M$ is equivalent to $\Vect$ so that $\Irr(\mc M)$ contains a single element, 2-injectivity boils down to the ordinary injectivity condition.
The second condition in definition \ref{def:2injectivity}, i.e. the spanning condition, is a form of \emph{MPO-injectivity}, where the MPO simply imposes that any two adjacent legs of the tensor carry the same label for the region between them. Moreover, in some cases, this condition readily agrees with the semi-injectivity condition of ref.~\cite{Molnar:2018hls}, though a general comparison of the two conditions is not straightforward. Indeed, while semi-injectivity was tailored to capture states such as the AKLT state on the square lattice, ours is rather meant to encode SPT phases for \emph{internal} symmetries that go beyond 0-form invertible ones. 

Although we focus on uniform tensor network representations with open boundary conditions, this manuscript and the proofs therein should be regarded as a template for various fundamental theorems. For instance, one could consider closed boundary conditions, in which case we may need to require the 2-injectivity condition spelt out above. A second variant concerns \emph{non-uniform} tensor network representations, whose tensors may vary from site to site. The latter may be required to recover the aforementioned classification of (2+1)d SPTs.

\bigskip\noindent
\textbf{Acknowledgements.} CD would like to thank Laurens Lootens and Frank Verstraete for comments. CD is grateful to Kansei Inamura and Alex Turzillo for illuminating exchanges at an early stage of this project. BK would like to thank Albert H. Werner for useful discussions. Large Language Models were used for proofreading and formulation improvements during the drafting process. The authors would like to thank the Isaac Newton Institute for Mathematical Sciences, Cambridge, for support and hospitality during the programme \emph{Mathematics of many-body entanglement}, where work on this paper was undertaken. This work was supported by EPSRC grant EP/Z000580/1. This work was partially supported by a grant from the Simons Foundation awarded to CD.
\bigskip \bigskip
\section{One dimension}

\subsection{Tensor network representations}

Let $d$ be a positive integer. We denote the standard basis of $\mathbb C^{\oplus d}$ by $\{e_i\}_{i \in [d]}$, where $[d] = \{1,\ldots,d\}$. Write $F^d$ for the \emph{free algebra} on the vector space $\mathbb C^{\oplus d}$, i.e. the vector space $\bigoplus_{L=0}^{\infty} (\mathbb C^{\oplus d})^{\otimes L}$ equipped with a multiplication rule given by the tensor product. We define a \emph{word} of length $L$ as a function $\fr i \colon \{1,\ldots,L\} \to [d]$. The set of words of length $L$ is denoted by $[d]^L$ and the set of all words together with concatenation forms the \emph{free monoid} $[d]^* = \bigsqcup_{L \geq 0}[d]^L$. Given a word $\fr i \equiv i_1 \, i_2 \, \cdots \, i_L \in [d]^L$, we write $e_\fr i \equiv e_{i_1} \otimes e_{i_2} \otimes \cdots \otimes e_{i_L}$ for the corresponding basis vector of $F^d$. We are interested in the tensor network states on $L$ sites, for any $L$, that live in the (physical) tensor product Hilbert space $(\mathbb C^{\oplus d})^{\otimes L} \subset F^d$.

\begin{definition}[Tensor network representation] 
    We define a tensor network representation on a chain with local Hilbert space $\mathbb C^{\oplus d}$ as a finite-dimensional module over the free algebra $F^d$.
\end{definition}

\noindent
Let us unpack the above definition. Let $V^A$ be a finite-dimensional complex vector space and $\rho^A \colon F^d \to \End_\mathbb C(V^A)$ be an algebra homomorphism. Since $F^d$ is freely generated by $\mathbb C^{\oplus d}$, the $F^d$-module structure on $V^A$ is fully specified by the $\mathbb C$-linear maps
\begin{equation}
    A^i \equiv \rho^A(e_i) \in \End_\mathbb C(V^A) \, ,
\end{equation}
for every $i \in [d]$. More generally, for every $\fr i \equiv i_1 \, i_2 \, \cdots \, i_L \in [d]^L$, we write $A^\fr i \equiv \rho^A(e_\fr i) = A^{i_1} \circ A^{i_2} \circ \cdots \circ A^{i_L} \in \End_\mathbb C(V^A)$. As customary, one interprets the map $e_i \mapsto A^i$ as a 3-valent tensor and draws
\begin{equation}
    A^i \equiv \MPS{A}{i} 
\end{equation}
so that $A^\fr i$, for every $\fr i \in [d]^*$, is obtained as a tensor contraction. Here, vector spaces are depicted as lines and $\mathbb C$-linear maps as coupons joining lines. 
The vector space $V^A$ is the \emph{bond space} of the tensor, and its dimension $D := \dim_\mathbb C V^A$ is referred to as its \emph{bond dimension}. Whenever convenient, the reader may identify $V^A$ with $\mathbb C^{\oplus D}$ and $\End_\mathbb C(V^A)$ with $\Mat_D(\mathbb C)$.

\subsection{Tensor network states} 

Let $V^A$ be an $F^d$-module with algebra homomorphism $\rho^A \colon F^d \to \End_\mathbb C(V^A)$. It encodes a tensor network representation on the chain. In the following, we are interested in tensor network states with open boundary conditions as specified by pairs $(v^A,\phi^A) \in V^A \times {V^A}^*$. Throughout, we treat vectors in $V^A$ as $\mathbb C$-linear maps $\mathbb C \to V^A$.

\begin{definition}[Uniform tensor network state]
    Given an $F^d$-module $V^A$ and a choice of boundary data $(v^A,\phi^A)$, we assign to every length $L$ the state
    \begin{equation} 
        \sum_{\fr i \in [d]^L} \big[ \phi^A \circ A^\fr i \circ  v^A \big] \, e_{\fr i} \in (\mathbb C^{\oplus d})^{\otimes L} \, .
    \end{equation}
\end{definition}

\noindent
In the literature, tensor network states of this form are referred to as (uniform) \emph{matrix product states} \cite{ACCARDI1981169,PhysRevLett.59.799,Affleck1988,Fannes1992}. 
In practice, one rarely deals with arbitrary matrix product states, and instead assume that the defining tensor---or rather the algebra homomorphism $\rho^A$---satisfies various properties. These properties may be physically or mathematically motivated. Here, since our priority is to prove a fundamental theorem in a way that generalises to higher dimensions, we are primarily interested in making assumptions that are as weak as possible while facilitating the mathematical derivations. 
Concretely, we are interested in tensor network states associated with a special class of boundary conditions:

\begin{definition}[(Co)cyclicity]
    Given an $F^d$-module $V^A$, we say that the boundary vector $v^A$ is `cyclic' if
    \begin{equation}
        \Span_\mathbb C\big\{A^\fr i \circ v^A \, \big| \, \fr i \in [d]^*\big\} = V^A \, ,
    \end{equation}
    and that the boundary covector $\phi^A$ is `cocyclic' if 
    \begin{equation}
        \Span_\mathbb C \big\{\phi^A \circ A^\fr i \, \big| \, \fr i \in [d]^*\big\} = {V^A}^* \, .
    \end{equation}
\end{definition}

\noindent
Equivalently, $v^A$ is cyclic if $\rho^A(-) \circ v^A \colon F^d \to V^A$ is surjective, and $\phi^A$ is cocyclic if $\phi^A \circ \rho^A(-) \colon F^d \to {V^A}^*$ is surjective.

Let us clarify how our assumption fits within the landscape of assumptions that are typically made. The most common one is to ask is to ask for $\Span_\mathbb C \{A^i \, | \, i \in [d]\} = \End_\mathbb C(V^A)$, without reference to any boundary conditions. Such a tensor network representation is said to be \emph{injective}. The terminology refers to the injectivity of the dual map $\End(V^A)^* \to (\mathbb C^{\oplus d})^*$, which is equivalent to the surjectivity of $e_i \mapsto A^i$, $i \in [d]$. A weaker version requires the existence of a length $L$ such that $\Span_\mathbb C \{ A^\fr i \, | \, \fr i \in [d]^L\} = \End_\mathbb C(V^A)$. Such tensor network representations that become injective after blocking $L$ tensors together are said to be \emph{normal}, and the smallest such $L$ is known as the \emph{injectivity length}. A yet weaker version requires $\Span_\mathbb C \{A^i \, | \, \fr i \in [d]^*\} = \End_\mathbb C(V^A)$, which allows for superpositions of products of arbitrary lengths. Equivalently, $\rho^A$ is surjective, which is equivalent to the $F^d$-module $V^A$ being simple by Burnside's theorem. Surjectivity of $\rho^A$ implies that any $v^A \neq 0$ is cyclic and any $\phi^A \neq 0$ is cocyclic. Our assumption only requires the chosen boundary conditions to be (co)cyclic. We adopt it since it comes at no extra cost here. Note that in two dimensions, for reasons that will be clarified in sec.~\ref{sec:PEPS_states}, we will employ a notion of (co)cyclicity involving a single tensor, whose one-dimensional analogue would read $\Span_\mathbb C \{A^i \circ v^A \, | \, i \in [d]\} = V^A$ and $\Span_\mathbb C \{\phi^A \circ A^i \, | \, i \in [d]\} = {V^A}^*$.  

We are now ready to state our running hypothesis.

\begin{hypothesis}\label{hyp:MPS}
    Let $V^A$ and $V^B$ be two $F^d$-modules such that $\dim_\mathbb C V^B \leq \dim_\mathbb C V^A$. Moreover, let $(v^A,\phi^A)$ be boundary data such that $v^A$ is cyclic and $\phi^A$ is cocyclic. We suppose the existence of boundary data $(v^B,\phi^B)$ such that
    \begin{equation}
        \label{eq:MPS_hypothesis}
        \phi^A \circ A^\fr i \circ v^A = \phi^B \circ B^\fr i \circ v^B \, ,
    \end{equation}
    for every $\fr i \in [d]^*$.
\end{hypothesis}

\noindent
Specifically, on a chain of length $L=3$, hypothesis \ref{hyp:MPS} stipulates that
\begin{equation}
    \MPSthree[A]{i_1}{i_2}{i_3}
    =
    \MPSthree[B]{i_1}{i_2}{i_3}
\end{equation}
for every $i_1,i_2,i_3 \in [d]$.

\bigskip 

\subsection{Fundamental theorem}

Hypothesis \ref{hyp:MPS} does not make any assumptions about the $F^d$-module $V^B$ and the boundary data $(v^B,\phi^B)$. The following lemma shows that such assumptions are not required as they follow from eq.~\eqref{eq:MPS_hypothesis} alone. 

\begin{lemma}\label{lemma:MPS_boundary}
    Under hypothesis \ref{hyp:MPS}, $v^B$ is a cyclic vector and $\phi^B$ is a cocyclic covector.
\end{lemma}
\noindent
\begin{proof}
    For every $\fr i, \fr i' \in [d]^*$, regard
    \begin{equation}
        \label{eq:MPS_pairing}
        (e_{\fr i'},e_{\fr i}) \mapsto \phi^A \circ A^{\fr i' \, \fr i} \circ v^A = (\phi^A \circ A^{\fr i'}) \circ (A^\fr i \circ v^A) 
    \end{equation}
    as a bilinear form on $F^d \times F^d$. This form pairs the vectors $A^{\fr i} \circ v^A$, $\fr i \in [d]^*$, which span $V^A$ by cyclicity, with the covectors $\phi^A \circ A^{\fr i'}$, $\fr i' \in [d]^*$, which span ${V^A}^*$ by cocyclicity. Therefore, the rank of this map equals $\dim_\mathbb C V^A$. By hypothesis \ref{hyp:MPS}, the form obtained by replacing $A$ by $B$ possesses the same rank. But the same factorisation shows that it factors through $\Span_\mathbb C\{B^\fr i \circ v^B \, | \, \fr i \in [d]^*\}$, so that
    \begin{equation}
        \dim_\mathbb C V^A \leq \dim_\mathbb C \Span_\mathbb C \big\{ B^\fr i \circ v^B  \, \big| \, \fr i \in [d]^*\big\} \leq \dim_\mathbb C V^B \, .
    \end{equation}
    Since $\dim_\mathbb C V^B \leq \dim_\mathbb C V^A$ by hypothesis \ref{hyp:MPS}, this forces $\Span_\mathbb C \{B^\fr i \circ v^B \, | \, \fr i \in [d]^*\} = V^B$, i.e. $v^B$ cyclic. Mirroring the argument yields $\phi^B$ cocyclic. 
\end{proof}

\begin{theorem}[Fundamental theorem\footnote{In spite of the name, `fundamental theorem' does not refer to a single theorem, but rather to a family of theorems parametrised by three variables. Firstly, the assumptions made about the tensor themselves; here, we only assume the boundary data to be (co)cyclic. Secondly, the nature of the boundary conditions, which are most often taken to be closed and periodic. In this case, one would have required the algebra generated by the matrices $A^i$ to be semisimple; a proof of the corresponding fundamental theorem can be found in disguise in ref.~\cite{MR1610851}. Thirdly, the hypothesis itself: whether the states are required to be equal or merely proportional, and whether for all lengths or only for a fixed length determined by the injectivity length. One could just as easily have formulated our hypothesis for a fixed length, but we believe that the additional technical complications would have been distracting in the higher-dimensional case. }]\label{th:MPS_fundamental}
    Under hypothesis \ref{hyp:MPS}, there is a unique isomorphism of $F^d$-modules $Q \colon V^A \to V^B$ such that $Q \circ v^A = v^B$ and $\phi^B \circ Q = \phi^A$.  
\end{theorem}
\noindent
\begin{proof}
    By cocyclicity, $\phi^A \circ \rho^A(-) \colon F^d \to {V^A}^*$ is surjective. Therefore, for every $x \in F^d$, $x \in \ker [\rho^A(-) \circ v^A]$ if and only if $(\phi^A \circ A^{\fr i}) \circ (\rho^A(x) \circ v^A) =0$, for every $\fr i \in [d]^*$. By lemma \ref{lemma:MPS_boundary}, the same holds with $A$ replaced by $B$. But, under hypothesis \ref{hyp:MPS},
    \begin{equation}
    \begin{split}
        (\phi^B \circ B^{\fr i}) \circ (\rho^B(x) \circ v^B) 
        = (\phi^A \circ A^{\fr i}) \circ (\rho^A(x) \circ v^A) \, , 
    \end{split}
    \end{equation}
    which implies that $\ker[\rho^A(-) \circ v^A] = \ker [\rho^B(-) \circ v^B]$. By cyclicity of $v^A$ and lemma \ref{lemma:MPS_boundary}, the morphisms of $F^d$-modules $\rho^A(-) \circ v^A \colon F^d \to V^A$ and $\rho^B(-) \circ v^B \colon F^d \to V^B$ are surjective.  By the first isomorphism theorem, we finally obtain
    \begin{equation}
        V^A \cong F^d / \ker [\rho^A(-) \circ v^A] = F^d / \ker [\rho^B(-) \circ v^B] \cong V^B \, .
    \end{equation}
    Explicitly, this isomorphism is given by $Q \colon V^A \to V^B$, $A^\fr i \circ v^A \mapsto B^\fr i \circ v^B$, for every $\fr i \in [d]^*$. In particular $Q \circ v^A = v^B$, and $\phi^B \circ Q = \phi^A$ by hypothesis \ref{hyp:MPS}. Since $v^A$ is cyclic, there is a unique morphism of $F^d$-modules satisfying $Q \circ v^A = v^B$.
\end{proof}

\medskip \noindent
Since $F^d$ is generated by $\mathbb C^{\oplus d}$, an isomorphism of $F^d$-modules $Q \colon V^A \to V^B$ is precisely an invertible $\mathbb C$-linear map such that $Q \circ A^i = B^i \circ Q$, for every $i \in [d]$; this is the form in which the fundamental theorem typically appears. Graphically,
\begin{equation}
    \MPSQ{i}{l} = \MPSQ[B]{i}{r} \, ,
\end{equation}
for every $i \in [d]$. 
Let us emphasise that the above proof differs significantly from the usual ones \cite{Perez-Garcia:2006nqo,PhysRevLett.100.167202,Cuevas:2017jlr}. Purely algebraic, our proof is particularly concise as it exploits the first isomorphism theorem. Unfortunately, the latter does not not admit an immediate categorification that could be exploited in the setting of the next section. It is therefore useful to construct the isomorphism $Q$ more explicitly.

By cyclicity, $\Span_\mathbb C \{A^\fr i \circ v^A \, | \, \fr i \in [d]^*\} \cong V^A$. Therefore, the assignment $A^\fr i \circ v^A \mapsto B^\fr i \circ v^B$ defines $Q$ everywhere, provided it is well defined. Suppose $\sum_{\fr i \in [d]^*} c_\fr i \, ( A^\fr i \circ v^A) = 0$ for some coefficients $c_\fr i \in \mathbb C$. By hypothesis \ref{hyp:MPS}, 
\begin{equation}
    (\phi^B \circ B^{\fr i'}) \circ \bigg( \sum_{\fr i \in [d]^*} c_\fr i \, ( B^\fr i \circ v^B) \bigg)
    = (\phi^A \circ A^{\fr i'}) \circ \bigg( \sum_{\fr i \in [d]^*} c_\fr i \, ( A^\fr i \circ v^A) \bigg) = 0 \, ,
\end{equation}
for every $\fr i' \in [d]^*$. By lemma \ref{lemma:MPS_boundary}, we have $\Span_\mathbb C \{\phi^B \circ B^{\fr i'} \, | \, \fr i' \in [d]^*\} = {V^B}^*$. Hence $\sum_{\fr i \in [d]^*}c_\fr i \, ( B^\fr i \circ v^B) = 0$, which makes $Q$ well defined. Exchanging the roles of $A$ and $B$ produces a well-defined map $V^B \to V^A$, $B^\fr i \circ v^B \mapsto A^\fr i \circ v^A$, which is the inverse of $Q$. Moreover, 
\begin{equation}
\begin{split}
    Q \circ A^{i} \circ (A^\fr i \circ v^A) =
    Q \circ (A^{i \, \fr i} \circ v^A) = B^{i \, \fr i} \circ v^B = B^{i} \circ B^{\fr i} \circ v^B = B^{i} \circ Q \circ (A^{\fr i} \circ v^A) \, ,
\end{split}
\end{equation}
for every $i \in [d]$ and $\fr i \in [d]^*$. 
By cyclicity, we conclude that $Q \circ A^i = B^i \circ Q$, for every $i \in [d]$. Finally, choosing $\fr i = \emptyset$ yields $Q \circ v^A = v^B$, while hypothesis \ref{hyp:MPS} yields
\begin{equation}
    \phi^B \circ Q \circ (A^\fr i \circ v^A) = \phi^B \circ B^\fr i \circ v^B = \phi^A \circ A^\fr i \circ v^A \, ,
\end{equation}
for every $\fr i \in [d]^*$, so that $\phi^B \circ Q = \phi^A$ by cyclicity. 

\bigskip \noindent
This concludes our treatment of the one-dimensional case. Building on the algebraic formulation presented in this section, we introduce in the next section a new notion of two-dimensional tensor network representation, together with a corresponding set of assumptions, and a proof of the associated fundamental theorem.
\bigskip \bigskip
\section{Two dimensions}

\subsection{Categories of physical Hilbert spaces}

Let $d$ be a positive integer. Our first task is to construct analogues of the free algebra $F^d$, which will provide the (physical) tensor product Hilbert spaces---with local factor $\mathbb C^{\oplus d} = \Span_\mathbb C\{e_1,\ldots,e_d\}$---where our two-dimensional tensor network states reside. The difficulty here is two-fold. On the one hand, the dimensionality of the physical system dictates that two algebraic rules are required in order to generate Hilbert spaces that extend both in the vertical and horizontal directions; this necessitates a mathematical object with a higher level of abstraction, which we choose to model as a monoidal category. On the other hand, we wish to consider tensor network states for a fixed lattice geometry (e.g. square, hexagonal or triangular); a choice that must be explicitly encoded into this monoidal category. 

Our solution is to consider the free monoidal category over a monoidal signature, where the monoidal signature encodes both the choice of lattice geometry and the dimension of the local Hilbert space: 

\begin{definition}[Monoidal signature \cite{JOYAL199155}]
    A monoidal signature is a quadruple $\Sigma \equiv (\Sigma_0,\Sigma_1,{\rm s}, {\rm t})$ consisting of a set $\Sigma_0$ of generating objects, a set $\Sigma_1$ of generating morphisms, and two functions ${\rm s}, {\rm t} \colon \Sigma_1 \to (\Sigma_0)^* = \bigsqcup_{L \ge 0}\Sigma_0^L$, referred to as the source and target maps, respectively. 
\end{definition}

\noindent
A homomorphism of monoidal signatures between two monoidal signatures consists of two maps between their respective sets of generating objects and between their respective sets of generating morphisms that are compatible with the source and target maps. The category of monoidal signatures and homomorphisms of monoidal signatures is denoted by $\msf{MonSign}$. There is a forgetful functor ${\rm Forg} \colon \msf{StrMonCat} \to \msf{MonSign}$ between the category of (small) strict monoidal categories and the category of monoidal signatures obtained by stripping a strict monoidal category of its identity object, identity morphisms, tensor product and composition rule. 

\begin{definition}[Free strict monoidal category \cite{JOYAL199155,Selinger2011}] 
    Given a monoidal signature $\Sigma$, the free strict monoidal category $\mc F(\Sigma)$ on $\Sigma$ is obtained by applying the left adjoint of the forgetful functor ${\rm Forg} \colon \msf{StrMonCat} \to \msf{MonSign}$ to $\Sigma$.\footnote{Recall that the free algebra $F(V)$ on a vector space $V$ is obtained by applying the left adjoint of the forgetful functor, which sends an associative algebra to its underlying vector space, to $V$.}
    The $\mathbb C$-linearisation of $\mc F(\Sigma)$---which we also denote by $\mc F(\Sigma)$---is the $\mathbb C$-linear category with the same objects as $\mc F(\Sigma)$, and whose hom-spaces are the free $\mathbb C$-linear modules over the hom-sets of $\mc F(\Sigma)$. 
\end{definition}

\noindent
Concretely, objects in the free strict monoidal category $\mc F(\Sigma)$ are words in the free monoid $(\Sigma_0)^*$; morphisms are obtained by composing and tensoring generating morphisms from $\Sigma_1$ along with identity morphisms, subject to the bifunctoriality of the tensor product. Graphically, morphisms in $\mathcal{F}(\Sigma)$ are represented by string diagrams with wires and boxes labelled by elements of $\Sigma_0$ and $\Sigma_1$, respectively; these diagrams are constructed to be compatible with the source and target maps, and are considered up to planar isotopies that keep every wire monotone with respect to the `time direction'.

\bigskip \noindent
We are now ready to construct our main example. Consider the monoidal signature whose set of generating objects is $\{X_1,X_2\}$ and whose set of generating morphisms is $\{f_i \, | \, i \in [d]\}$, where ${\rm s}(f_i) = X_1X_2 \equiv X_1 \otimes X_2$ and ${\rm t}(f_i) = X_2X_1 \equiv X_2 \otimes X_1 \in \{X_1,X_2\}^*$, for every $i \in [d]$.  We denote the free (strict $\mathbb C$-linear) monoidal category on this signature by $\Fsq$. 

By definition, objects in $\Fsq$ are words $\fr X$ in the free monoid $\{X_1,X_2\}^*$. The hom-space from the word $\fr X \in \{X_1,X_2\}^*$ to the word $\fr X' \in \{X_1,X_2\}^*$ is non-zero  if and only if $\fr X'$ can be obtained from $\fr X$ by replacing instances of the factor $X_1X_2$ with $X_2 X_1$. If $\fr X'$ can be obtained from $\fr X$ by such adjacent rewrites, we say that $\fr X'$ is `reachable' from $\fr X$, or that $\fr X$ `reaches' $\fr X'$, and write $\fr X \preceq \fr X'$. The hom-space of morphisms from $\fr X$ to $\fr X'$ is then isomorphic to $(\mathbb C^{\oplus d})^{\otimes N}$, where $N$ is the number of such rewrites, which does not depend on the order in which they are made. 

In the following, we are especially interested in `connected' hom-spaces 
between two words $\fr X \in \{X_1,X_2\}^*$ and $\fr X' \succeq \fr X$  such that none of the non-zero morphisms in this hom-space can be factored as a non-trivial product of morphisms. This implies, in particular, that the first and last letters of  $\fr X$ are $X_1$ and $X_2$, respectively, whereas the first and last letters of $\fr X'$ are $X_2$ and $X_1$, respectively. For instance, choose $\fr X := X_1 X_1 X_1 X_2 X_1 X_2 X_2$ and $\fr X' := X_2 X_1 X_2 X_1 X_2 X_1 X_1$. The string diagram representing any basis morphism in this hom-space is of the form\footnote{String diagrams representing morphisms in a hom-space that is not connected always contain individual wires and/or disconnected pieces, hence the terminology.}
\begin{equation}
    \label{eq:stringDiag_square}
    \stringDiag{1} \rightsquigarrow \stringDiag{2} \, .
\end{equation}
We obtained the diagram on the r.h.s. of eq.~\eqref{eq:stringDiag_square} by applying a planar isotopy so as to reveal the underlying square geometry. 
In particular, given a choice of $\fr X$ and reachable $\fr X'$ from $\fr X$, the topology of all string diagrams is fixed. Generally, any connected hom-space can be assigned the geometry of an open patch of the square lattice, which is obtained by cutting out a disc-like region whose boundary is intersected transversely by the edges of the lattice. In other words, every connected hom-space of $\Fsq$ is of the form $(\mathbb C^{\oplus d})^{\otimes N}$, where $N$ is the number of vertices in some open patch of the square lattice. Notice that not every open patch of the square lattice can be obtained in this way. A convenient way to express the accessible shapes is in terms of connected skew Young diagrams:

\begin{definition}[Young diagram]
    The Young diagram corresponding to a finite sequence $\lambda_1 \geq \lambda_2 \geq \cdots \geq \lambda_L \geq 0$ of natural numbers consists of $L$ left-justified rows of boxes stacked from top to bottom, containing $\lambda_1, \lambda_2, \ldots, \lambda_L$ boxes, respectively. 
\end{definition}

\begin{definition}[Connected skew Young diagram]
    The skew Young diagram corresponding to two finite sequences $\lambda \equiv (\lambda_1,\lambda_2,\ldots,\lambda_L)$ and $\lambda' \equiv (\lambda_1',\lambda_2',\ldots,\lambda_L')$ of natural numbers such that $\lambda_1 \geq \lambda_1' , \lambda_2 \geq \lambda_2', \ldots, \lambda_{L} \geq \lambda_{L}'$ consists of the Young diagram corresponding to the sequence $\lambda$, from which the boxes of the Young diagram corresponding to the sequence $\lambda'$ have been removed, assuming both diagrams are top-left justified. We refer to the resulting diagram as $\nicefrac{\lambda}{\lambda'}$ with the convention that zeros are dropped from the finite sequences.  A skew Young diagram is said to be connected if every two boxes can be joined by a chain of boxes sharing at least one common edge. We denote the set of all connected skew Young diagrams by $\Yc$.
\end{definition}

\noindent 
For instance, consider the finite sequences $4 \geq 4 \geq 3 \geq 0 $ and $2 \geq 1 \geq 0 \geq 0$. The corresponding connected skew Young diagram $\nicefrac{(4,4,3)}{(2,1)}$ is given by
\begin{equation}
    \young{1} \; \equiv \young{2} \, ,
\end{equation}
which we recognise as the Poincaré dual of the graph underlying the string diagram on the r.h.s. of eq.~\eqref{eq:stringDiag_square}. More generally, the following statement holds:

\begin{proposition}
    Connected hom-spaces in $\Fsq$ are in one-to-one correspondence with connected skew Young diagrams in $\Yc$ up to translation. 
\end{proposition}

\noindent
\begin{proof}
    Read from the bottom-left to the top-right, every $Y \in \Yc$ specifies a connected hom-space in $
    \Fsq$, whereby each horizontal edge corresponds to a wire labelled by $X_1$, each vertical edge to a wire labelled by $X_2$, and each box to a copy of $\Hom_{\Fsq}(X_1 X_2,X_2 X_1) \cong \mathbb C^{\oplus d}$. Conversely, let $\fr X$ and $\fr X'$ be two words in $\{X_1,X_2\}^*$ such that $\fr X \preceq \fr X'$ and the corresponding hom-space is connected. Define the finite sequence $\lambda(\fr X)$ by letting $\lambda(\fr X)_i$ be the number of letters $X_1$ to the left of the $i$-th letter $X_2$ counted from the right. Define $\lambda(\fr X')$ similarly. One can verify that $\lambda(\fr X)_i \geq \lambda(\fr X')_i$, for every $i$, so that we obtain a connected skew Young diagram, namely $\nicefrac{\lambda(\fr X)}{\lambda(\fr X')}$. These two constructions are clearly inverse to each other.
\end{proof}

\medskip
\noindent
Henceforth, we refer to a
connected skew Young diagram in $\Yc$ simply as a `shape', and label a connected hom-space via its corresponding shape. Given a shape $Y \in \Yc$, we denote by ${\rm s}(Y)$ and ${\rm t}(Y)$ the source and target objects in $\Fsq$, respectively, of the morphisms in the corresponding connected hom-space.
Moreover, we write $[d]^Y$ for the set of functions $\fr i \colon Y \to [d]$ and $[d]^{\Yc} := \bigsqcup_{Y \in \Yc}[d]^{Y}$. We identify $\Hom_{\Fsq}(X_1X_2,X_2X_1) = \Span_\mathbb C\{f_i \, | \, i \in [d]\}$ with $\mathbb C^{\oplus d}$ via $f_i \mapsto e_i$.
Finally, for $\fr i \in [d]^Y$, we denote by $e_\fr i \in \Hom_{\Fsq}({\rm s}(Y), {\rm t}(Y))$ the morphism obtained by placing at each box of $Y$ the basis vector in $\mathbb C^{\oplus d}$ labelled by the value of $\fr i$ on that box.

\bigskip \noindent
A similar strategy produces tensor product Hilbert spaces over the vertices of some open patch of the hexagonal lattice. Consider the monoidal signature whose set of generating objects is $\{X_1,X_2,X_3\}$ and whose set of generating morphisms is $\{f_{i_1}, f'_{i_2} \, | \, i_1,i_2 \in [d]\}$, where ${\rm s}(f_{i_1}) = X_1$, ${\rm t}(f_{i_1}) = X_2 X_3$, ${\rm s}(f'_{i_2}) = X_3X_2$ and ${\rm t}(f'_{i_2}) = X_1$, for every $i_1,i_2 \in [d]$. We denote the free monoidal category on this signature by $\Fhex$, which can be analysed in the same vein as the previous example. 

\bigskip

\subsection{2-vector spaces}

Above, we introduced an algebraic structure in the form of a monoidal category in order to encode physical tensor product Hilbert spaces associated with square geometries of arbitrary size. While the same could have been accomplished via a more pedestrian approach, the merit of our formalism is that it provides a particularly elegant and basis-independent    way of defining two-dimensional tensor network representations, and of concisely expressing some of their key properties. 

The first step towards defining our two-dimensional tensor network representation is to introduce a categorified notion of bond space, which we choose to model as a 2-vector space:\footnote{There exist numerous alternative notions of 2-vector spaces; see ref.~\cite{BAEZ1997125} and references therein.}

\begin{definition}[2-vector space]
    We define a 2-vector space as a $\mathbb C$-linear, finite, semisimple category. The 2-category of 2-vector spaces, $\mathbb C$-linear functors and natural transformations is denoted by $\TVect$.
\end{definition}

\noindent 
Suppose $\mc M = \Vect^{\boxplus D}$ for some positive integer $D$. An object in $\mc M$ can be thought of as a $D$-dimensional $\Vect$-valued vector, and a morphism as a $D$-dimensional vector of $\mathbb C$-linear maps acting entrywise \cite{kapranov19942}. Similarly, a $\mathbb C$-linear endofunctor of $\mc M$ can be thought as a $D \times D$ $\Vect$-valued matrix acting via matrix multiplication with $(\otimes,\oplus)$ in lieu of $(\times,+)$. Every 2-vector space of rank $D$ is equivalent to $\Vect^{\boxplus D}$. In what follows, the reader may therefore identify $\mc M$ with $\Vect^{\boxplus D}$ whenever convenient.

Let $\mc M$ be a 2-vector space and $\Irr(\mc M)$ a set of representatives of isomorphism classes of simple objects. A $\mathbb C$-linear endofunctor $T \colon \mc M \to \mc M$ is determined, up to natural isomorphism, by its action on simple objects. Writing $T^{M_2}_{M_1} := \Hom_\mc M (M_2,T(M_1))$, there is for every $M_1 \in \Irr(\mc M)$ a canonical isomorphism 
\begin{equation}
    T(M_1) \cong  \bigoplus_{M_2 \in \Irr(\mc M)} T^{M_2}_{M_1} \otimes M_2\, .
\end{equation}
For every object $M$ in $\mc M$, we then have
\begin{equation}
    T(M) \cong \bigoplus_{M_1 \in \Irr(\mc M)} \Hom_\mc M(M_1,M) \otimes T(M_1) \, ,
\end{equation}
with the action on morphisms given by postcomposition on the first $\otimes$-factor. The composite $T_1 \circ T_2$ of two endofunctors $T_1,T_2 \colon \mc M \to \mc M$ is given by
\begin{equation}
    \label{eq:PEPS_compFunc}
    (T_1 \circ T_2)^{M_3}_{M_1} \cong \bigoplus_{M_2 \in \Irr(\mc M)} (T_1)^{M_3}_{M_2} \otimes (T_2)^{M_2}_{M_1}  \, ,
\end{equation}
for every $M_1,M_3 \in \Irr(\mc M)$.
A natural transformation $\eta \colon T_1 \Rightarrow T_2$ amounts to a family of morphisms $\eta_{M_1} \colon T_1(M_1) \to T_2(M_1)$, $M_1 \in \Irr(\mc M)$. 
By semisimplicity, 
\begin{equation}
    \Hom_\mc M \big(T_1(M_1), T_2(M_1)\big) \cong \bigoplus_{M_2 \in \Irr(\mc M)} {(T_1)^{M_2}_{M_1}}^* \otimes (T_2)^{M_2}_{M_1} \, ,
\end{equation}
where we used $\Hom_\mc M\big(T_1(M_1),M_2\big) \cong \Hom_\mc M\big(M_2,T_1(M_1) \big)^*$ via the canonical pairing. Thus,
\begin{equation} 
    \label{eq:PEPS_NatBlockDecomp}
    \Nat(T_1, T_2) \cong 
    \!\!
    \bigoplus_{M_1, M_2 \in \Irr(\mc M)} \!\!
    {(T_1)^{M_2}_{M_1}}^* \otimes (T_2)^{M_2}_{M_1} \, .
\end{equation}
Defining the object $M_{\oplus} = \bigoplus_{M_1 \in \Irr(\mc M)} M_1$ and, for every $\mathbb C$-linear endofunctor $T$, the vector space
\begin{equation}
    \label{eq:PEPS_VT}
    V_T := \Hom_\mc M\big( M_\oplus,T(M_\oplus)\big) \cong \bigoplus_{M_1, M_2 \in \Irr(\mc M)} T_{M_1}^{M_2},
\end{equation}
every natural transformation $\eta \colon T_1 \Rightarrow T_2$ induces a $\mathbb C$-linear map 
\begin{equation}
    \label{eq:PEPS_NatInducedMap}
    \eta_{M_{\oplus}} \circ - \colon V_{T_1} \to V_{T_2} \, ,
\end{equation} 
which acts blockwise, mapping $(T_1)_{M_1}^{M_2}$ into $(T_2)_{M_1}^{M_2}$. 
By eq.~\eqref{eq:PEPS_NatBlockDecomp}, this identifies $\Nat(T_1,T_2)$ with the space of such block-diagonal maps. 
In the following, we regard these induced $\mathbb C$-linear maps as 2-valent tensors.

\bigskip \noindent
At this stage, it is convenient to introduce a graphical calculus, which we borrow from 2-category theory. Simple objects of $\mc M$ label planar regions, and the space $T_{M_1}^{M_2}$ is depicted as an oriented line at the interface of two regions labelled by $M_1$ and $M_2$, respectively. Leaving the regions unlabelled then amounts to summing over all possible labels:
\begin{equation}
    \TLine{T}{}{}{}{} \equiv 
    \bigoplus_{M_1,M_2 \in \Irr(\mc M)} 
    \TLine{T}{M_1}{M_2}{}{} \, .
\end{equation}
A vector $v \in T_{M_1}^{M_2}$ and a covector $\phi \in {T_{M_1}^{M_2}}^*$ are depicted as coupons ending the line labelled by $T$:
\begin{equation}
    v \equiv \TLine{T}{M_1}{M_2}{v}{} \q \text{and} \q  \phi \equiv \TLine{T}{M_1}{M_2}{}{\phi} \, .
\end{equation}
Composition of endofunctors corresponds to stacking of lines, so that eq.~\eqref{eq:PEPS_compFunc} reads
\begin{equation}
    \label{eq:PEPS_compT}
    \TLine{T_1 \circ T_2}{M_1}{M_3}{}{} \equiv \bigoplus_{M_2 \in \Irr(\mc M)}\TTLine{T_1}{T_2}{M_1}{M_2}{M_3}{}{}{}{} \, .
\end{equation}
The $\mathbb C$-linear map $\eta_{M_\oplus} \circ -$ induced by any natural transformation $\eta \colon T_1 \Rightarrow T_2$ is depicted as a coupon joining two lines labelled by $T_1$ and $T_2$, respectively: 
\begin{equation}
    \label{eq:PEPS_coupon}
    \eta_{M_\oplus} \circ -  \equiv \NatLine{T_1}{T_2}{}{}{\eta} \, .
\end{equation}
Labelling the two regions by $M_1$  and $M_2$, respectively, amounts to selecting the block $(T_1)^{M_2}_{M_1} \to (T_2)^{M_2}_{M_1}$ (cf. eq.~\eqref{eq:PEPS_NatBlockDecomp}). Combining eqs.~\eqref{eq:PEPS_compT} and \eqref{eq:PEPS_coupon}, the map induced by a natural transformation $\eta \colon T_1 \circ T_2 \Rightarrow T_3$ is depicted as 

\begin{equation}
    \label{eq:PEPS_NatComp}
    \eta_{M_\oplus} \circ - \equiv 
    \NatCoup{T_1}{T_2}{T_3}{}{}{}{\eta} = \bigoplus_{M_1,M_2,M_3}
    \NatCoup{T_1}{T_2}{T_3}{M_1}{M_2}{M_3}{\eta} \, .
\end{equation}
Explicitly, 
\begin{equation}
    \label{eq:PEPS_NatTDecomp}
\begin{split}
   \Nat(T_1 \circ T_2, T_3 ) &\cong 
   \!\! \bigoplus_{M_1,M_3 \in \Irr(\mc M)} \!\!
   {(T_1 \circ T_2)^{M_3}_{M_1}}^* \otimes (T_3)^{M_3}_{M_1}
   \\ 
   &\cong  \bigoplus_{M_1,\ldots,M_3} {(T_2)^{M_2}_{M_1}}^* \otimes {(T_1)^{M_3}_{M_2}}^* \otimes (T_3)^{M_3}_{M_1}  \, ,
\end{split}
\end{equation}
so that labelling the regions in eq.~\eqref{eq:PEPS_NatComp} by $M_1,M_2,M_3 \in \Irr(\mc M)$ amounts to selecting the block $(T_1)_{M_2}^{M_3} \otimes (T_2)_{M_1}^{M_2} \to (T_3)_{M_1}^{M_3}$.

\bigskip

\subsection{Tensor network representations}
Recall that a one-dimensional tensor network representation is defined as a module structure on a bond space isomorphic to $\mathbb C^{\oplus D}$ over the free algebra $F^d$.
In direct parallel, two-dimensional tensor network representations are defined in terms of module categories over the free monoidal category. As is common in category theory, the main reason for introducing module categories is that one can then consider functors between them. Such functors will precisely be the output of our fundamental theorem.
Let us first introduce a categorification of the notion of module over an algebra:  

\begin{definition}[Strict module category \cite{Ostrik2003}] 
    We define a strict left module category over a $\mathbb C$-linear strict monoidal category $\mc C$ (or left $\mc C$-module category) as a 2-vector space $\mc M$ together with a $\mathbb C$-linear strict monoidal functor $\rho \colon \mc C \to \EndC(\mc M)$ into the strict monoidal category of $\mathbb C$-linear endofunctors of $\mc M$ and natural transformations.
\end{definition}
\noindent
For future reference, we introduce here module functors and module category equivalences.
\begin{definition}[Equivalence of module categories \cite{etingof2016tensor}] \label{def:PEPS_equivalence}
    Let $\mc M$ and $\mc M'$ be two left $\mc C$-module categories with strict monoidal functors $\rho \colon \mc C \to \End(\mc M)$ and $\rho' \colon \mc C \to \End(\mc M')$. A $\mc C$-module functor from $\mc M$ to $\mc M'$ is a pair $(\mc Q , \zeta)$ consisting of a $\mathbb C$-linear functor $\mc Q \colon \mc M \to \mc M'$ and a natural isomorphism $\zeta \colon \mc Q \circ \rho(-) \Rightarrow \rho'(-) \circ \mc Q$ with components $\zeta_{X,M} \colon \mc Q \big(\rho(X)(M) \big) \xrightarrow{\sim} \rho'(X)\big( \mc Q(M)\big)$, for every $X$ in $\mc C$ and $M$ in $\mc M$, such that $\zeta_{X \otimes X',M} = \rho'(X)(\zeta_{X',M}) \circ \zeta_{X,\rho(X')(M)}$. The $\mc C$-module functor $(\mc Q,\zeta)$ is a $\mc C$-module equivalence whenever $\mc Q$ is an equivalence of categories.
\end{definition}

\noindent
We are now ready to introduce our notion of two-dimensional tensor network representation:
\begin{definition}[Tensor network representation]
    We define a uniform tensor network representation on the square (resp. hexagonal) geometry with local Hilbert space $\mathbb C^{\oplus d}$ as a strict module category over the free strict $\mathbb C$-linear monoidal category $\Fsq$ (resp. $\Fhex$).
\end{definition}

\noindent
Let us unpack the above definition for the square geometry. 
Let $\mc M^A$ be a 2-vector space and $\rho^A \colon \Fsq \to \EndC(\mc M^A)$ be a $\mathbb C$-linear strict monoidal functor. 
For every word $\fr X \in \{X_1,X_2\}^*$, we write $T_\fr X^A \equiv \rho^A(\fr X)$ for the composite of copies of endofunctors $T_1^A \equiv T_{X_1}^A$ and $T_2^A \equiv T_{X_2}^A$, one for each letter of $\fr X$, and
\begin{equation}
    \label{eq:PEPS_VAX}
    V_{\fr X}^A \equiv V_{T_\fr X^A} = \!\! \bigoplus_{M_1,M_2 \in \Irr(\mc M^A)} \!\! (T_\fr X^A)_{M_1}^{M_2}
\end{equation}
for its associated vector space. 
Since $\Fsq$ is freely generated by the objects $X_1,X_2$ and morphisms $f_i \equiv e_i$, $i \in [d]$,  the $\Fsq$-module structure on $\mc M^A$ is fully specified by the endofunctors $T^A_1 := \rho^A(X_1)$ and $T^A_2 := \rho^A(X_2)$, together with natural transformations
\begin{equation}
    \rho^A(e_i) \in \Nat(T^A_1 \circ T^A_2, T^A_2 \circ T^A_1) = \Nat(T_{X_1X_2}^A,T_{X_2X_1}^A)\, .
\end{equation}
By eq.~\eqref{eq:PEPS_NatInducedMap}, each natural transformation $\rho^A(e_i)$ induces a $\mathbb C$-linear map
\begin{equation}
    A^i := \rho^A(e_i)_{M_\oplus} \circ - \colon V_{X_1X_2}^A \to V_{X_2X_1}^A \, .
\end{equation}
In the graphical calculus,\footnote{Notice that, for visual clarity,  we omitted the labels $T_1^A$ and $T_2^A$ on the vertical and horizontal lines, respectively. The little `physical leg' carrying the label $i$ is not strictly part of our graphical calculus. It should be thought of as coming out of the page, and it simply points to the blob, labelling it by $A^i$.}
\begin{equation}
    \label{eq:PEPS_T}
    A^i \equiv \PEPS{1}{A}{i}{}{}{}{}{}{}  = \bigoplus_{M_1,\ldots,M_4} \,  \PEPS{1}{A}{i}{M1234}{}{}{}{}{} \, .
\end{equation}
By eqs.~\eqref{eq:PEPS_compFunc} and \eqref{eq:PEPS_NatBlockDecomp},
\begin{equation}
    \label{eq:PEPS_NatBlockDecomp_T}
    \Nat(T_1^A \circ T_2^A, T_2^A \circ T_1^A) \cong \bigoplus_{M_1,\ldots,M_4} {(T_2^A)_{M_1}^{M_2}}^* \otimes {(T_1^A)_{M_2}^{M_4}}^* \otimes {(T_2^A)_{M_3}^{M_4}} \otimes {(T_1^A)_{M_1}^{M_3}} \, ,
\end{equation}
so that $A^i$ may be regarded as a 4-valent tensor in ${V^A_{X_2}}^* \otimes {V^A_{X_1}}^* \otimes V^A_{X_2} \otimes V^A_{X_1}$ which vanishes outside the blocks appearing in eq.~\eqref{eq:PEPS_NatBlockDecomp_T}. Labelling the regions in eq.~\eqref{eq:PEPS_T} by $M_1,\ldots, M_4 \in \Irr(\mc M^A)$ amounts to selecting the block $(T^A_1)^{M_4}_{M_2} \otimes (T_2^A)_{M_1}^{M_2} \to (T_2^A)_{M_3}^{M_4} \otimes (T_1^A)_{M_1}^{M_3}$. More generally, for every shape $Y\in \Yc$ and $\fr i \in [d]^Y$, we denote by 
\begin{equation}
    \label{eq:PEPS_Ai}
    A^\fr i \equiv \rho^A(e_\fr i)_{M_\oplus} \circ - \colon V_{{\rm s}(Y)}^A \to V_{{\rm t}(Y)}^A
\end{equation}
the $\mathbb C$-linear map induced by $\rho^A(e_\fr i) \in \Nat(T^A_{{\rm s}(Y)},T^A_{{\rm t}(Y)})$. 
By definition, $e_\fr i$ is a composite of morphisms of the form $\id_{\fr Y} \otimes e_i \otimes \id_{\fr Y'}$ for some words $\fr Y,\fr Y' \in \{X_1,X_2\}^*$, one for each box of $Y$. Accordingly, $A^\fr i$ can be written, in a non-unique way, as a composite of $\mathbb C$-linear maps induced by natural transformations of the form $\rho^A(\id_{\fr Y} \otimes e_i \otimes \id_{\fr Y'}) \colon T^A_{\fr Y X_1X_2 \fr Y'} \Rightarrow T^A_{\fr Y X_2X_1 \fr Y')}$. In other words, $A^\fr i$ is obtained by placing at each box of $Y$ the tensor $A^i$, with $i$ the value of $\fr i$ on that box, and contracting along edges of $Y$. Note that since the pairing along each edge is block-diagonal, the labels of the shared regions are automatically identified upon contraction.

\bigskip \noindent
To conclude this part, let us consider a couple of special cases. Suppose that $\mc M^A \simeq \Vect$. The endofunctors $T_1^A$ and $T_2^A$ then amount to complex vector spaces, which encode the bond degrees of freedom along the edges of a shape. As detailed in sec.~\ref{sec:PEPS_Vec}, this recovers the ordinary case of tensor network representations known as \emph{projected entangled pair states} (PEPS). The name refers to an equivalent description, whereby maximally entangled (EPR) states are placed on the edges, and at each site a $\mathbb C$-linear map from a tensor product of bond spaces to the physical space is applied. In this description, EPR pairs play the role of \emph{resource states}. Choosing different classes of resource states yields tensor network representations encoding distinct \emph{entanglement structures} \cite{Molnar:2018hls,Christandl:2020chy}. 
For instance, one may place GHZ states on the virtual degrees of freedom surrounding each plaquette.\footnote{Within our formalism, plaquettes of the tensor network are associated with vertices of the underlying shape.} Within our framework, this class of tensor network representations is obtained by choosing $\mc M^A = \Vect^{\boxplus D}$ and endofunctors $T_1^A, T_2^A$ such that $(T_1^A)_{M_1}^{M_2} \cong (T_2^A)_{M_1}^{M_2} \cong \mathbb C$, for all $M_1,M_2 \in \Irr(\mc M^A)$. 

\bigskip 

\subsection{Tensor network states} \label{sec:PEPS_states}
Let $\mc M^A$ be a strict $\Fsq$-module category with $\rho^A \colon \Fsq \to \EndC(\mc M^A)$. It encodes a tensor network representation on the square geometry. 
By eq.~\eqref{eq:PEPS_compFunc}, 
\begin{equation}
    \label{eq:PEPS_VAXX}
    V^A_{\fr X \fr X'} \cong 
    \bigoplus_{M_1,M_2,M_3} 
    \! (T^A_\fr X)_{M_2}^{M_3}
    \otimes (T^A_{\fr X'})_{M_1}^{M_2} \; \subseteq \,  V^A_{\fr X} \otimes V^A_{\fr X'} \, ,
\end{equation}
for every $\fr X,\fr X' \in \{X_1,X_2\}^*$. Furthermore, for every vector $v \in V_\fr X^A$, we write $v_{M_1}^{M_2} \in (T_\fr X^A)_{M_1}^{M_2}$ for its components, and similarly for covectors. Given $v \in V_\fr X^A$ and $v' \in V_{\fr X'}^A$, we define another vector $v \odot v' \in V^A_{\fr X \fr X'}$ by
\begin{equation}
    \label{eq:PEPS_odot}
    (v \odot v')_{M_1}^{M_3} := \sum_{M_2 \in \Irr(\mc M^A)} {v}_{M_2}^{M_3} \otimes {v'}_{M_1}^{M_2} \, ,
\end{equation}
i.e., the contraction of the two vectors along the label of the region they share. 
Given $\phi \in {V^A_\fr X}^*$ and $\phi' \in {V^A_{\fr X'}}^*$, we define $\phi \odot \phi'$ similarly. We denote the inclusion given in eq.~\eqref{eq:PEPS_VAXX} by $\iota^A \colon V^A_{\fr X \fr X'} \to V^A_{\fr X} \otimes V^A_{\fr X'}$, and let $\pi^A \colon V^A_{\fr X} \otimes V^A_{\fr X'} \to V^A_{\fr X \fr X'}$ be the projection $v \otimes v' \mapsto v \odot v'$, so that $\pi^A \circ \iota^A = \id$. Whenever convenient , we leave $\iota^A$ implicit and regard $V^A_{\fr X \fr X'}$ as a subspace of $V^A_\fr X \otimes V^A_{\fr X'}$.
Finally, we denote by $V_{\fr X}^A \odot V_{\fr X'}^A$ the subspace of $V_{\fr X}^A \otimes V_{\fr X'}^A$ spanned by the vectors $\iota^A \circ (v \odot v')$, so that $V^A_{\fr X \fr X'} \cong V^A_{\fr X} \odot V^A_{\fr X'}$ by eq.~\eqref{eq:PEPS_VAXX}. Besides, given a shape $Y \in \Yc$, $\fr i \in [d]^Y$ and words $\fr Y, \fr Y' \in \{X_1,X_2\}^*$, the $\mathbb C$-linear map induced by $\rho^A(\id_{\fr Y} \otimes e_\fr i \otimes \id_{\fr Y'})$ sends $v \odot v'' \odot v'$ to $v \odot (A^\fr i \circ v'') \odot v'$, for every $v \in V^A_\fr Y$, $v'' \in V^A_{{\rm s}(Y)}$ and $v' \in V^A_{\fr Y'}$. We denote this map by
\begin{equation}
    \label{eq:PEPS_idAid}
    \id_{V^A_{\fr Y}} \odot A^\fr i \odot \id_{V^A_{\fr Y'}} \colon V^A_{\fr Y {\rm s}(Y) \fr Y'} \to V^A_{\fr Y {\rm t}(Y) \fr Y'} \, .
\end{equation}
In order to construct states in the physical Hilbert spaces, one first needs to introduce boundary data.
We abbreviate 
\begin{equation}
    \label{eq:PEPS_abbreviation}
    V^A_{X_1} \equiv \VA1
    \, , \q
    V^A_{X_2} \equiv \VA2
    \, , \q
    V^A_{X_1X_2} \equiv V^A_{12}
    \, , \q
    V^A_{X_2X_1} \equiv V^A_{21} \, , \q  \ldots ,
\end{equation}
Choose boundary vectors $\vA1 \in \VA1$, $\vA2 \in \VA2$ and boundary covectors $\phA1 \in {\VA1}^*$, $\phA2 \in {\VA2}^*$. 
For every word $\fr X \in \{X_1,X_2\}^*$, we write $v_\fr X^A \in V^A_\fr X$ for the $\odot$-product of copies of $\vA1$ and $\vA2$, one for each letter of $\fr X$, i.e., the one-dimensional tensor network state obtained by contracting them
along the labels of the regions they share, and define $\phi^A_\fr X$ similarly. For instance, for the source and target of the shape $Y=(1)$,
\begin{align}
    v_{X_1 X_2}^A = \vA1 \odot \vA2 = \sum_{M_1,M_2,M_4} (\vA1)_{M_2}^{M_4} \otimes (\vA2)_{M_1}^{M_2} 
    \equiv
    \sum_{M_1,M_2,M_4}
    \TTLine{T^A_1}{T^A_2}{M_1}{M_2}{M_4}{\vA1}{\vA2}{}{}
    =
    \TTLine{T^A_1}{T^A_2}{}{}{}{\vA1}{\vA2}{}{}  ,
    \\
    \TTLine{T^A_2}{T^A_1}{}{}{}{}{}{\phi^A_2}{\phi^A_1} 
    = \sum_{M_1,M_3,M_4}
    \TTLine{T^A_2}{T^A_1}{M_1}{M_3}{M_4}{}{}{\phi^A_2}{\phi^A_1} 
    \equiv \sum_{M_1,M_3,M_4} (\phA2)_{M_3}^{M_4} \otimes (\phA1)_{M_1}^{M_3} = \phA2 \odot \phA1 = \phi_{X_2 X_1}^A \, .
\end{align}
In particular, $v^A_{X_1}\equiv \vA1$, $v^A_{X_2} \equiv \vA2$, $\phi^A_{X_1} \equiv \phA1$, $\phi^A_{X_2} \equiv \phA2$, and we abbreviate  $v_{X_1X_2}^A \equiv v_{12}^A$, $v_{X_2X_1}^A \equiv v^A_{21}$, $\phi^A_{X_2X_1} \equiv \phi^A_{21}$, etc. We are now ready to define tensor network states on the square geometry. 

\begin{definition}[Uniform tensor network state]
    Given a strict $\Fsq$-module category $\mc M^A$ and a choice of boundary data $(\vA1,\vA2,\phA1,\phA2)$, we assign to every shape $Y \in \Yc$ the state
    \begin{equation} 
        \sum_{\fr i \in [d]^Y} \big[ \phi^A_{{\rm t}(Y)} \circ A^\fr i \circ  v^A_{{\rm s} (Y)} \big] \, e_{\fr i} \in \Hom_{\Fsq} \big({\rm s}(Y), {\rm t}(Y) \big) \cong (\mathbb C^{\oplus d})^{\otimes |Y|} \, .
    \end{equation}
\end{definition}
\noindent
By definition, $V^A_{{\rm s}(Y)} = \bigoplus_{M_1,M_2} (T^A_{{\rm s}(Y)})_{M_1}^{M_2}$. For each summand, $M_1$ and $M_2$ label the two regions where the source and target boundaries of $Y$ meet, namely the top-right and bottom-left regions, henceforth referred to as the `corners' of $Y$. 
Since $A^\fr i$ is induced by a natural transformation, it maps the summand of $V^A_{{\rm s}(Y)}$ labelled by $M_1,M_2$ into the summand of $V^A_{{\rm t}(Y)}$ with the same labels. In other words, it preserves the labels of the two corners, so that
\begin{equation}
    \phi^A_{{\rm t}(Y)} \circ A^\fr i \circ  v^A_{{\rm s} (Y)}  = 
    \sum_{M_1,M_2}(\phi^A_{{\rm t}(Y)})^{M_2}_{M_1} \circ A^{\mathfrak i} \circ (v^A_{{\rm s}(Y)})^{M_2}_{M_1} \, ,
\end{equation}
for every $\fr i \in [d]^Y$.
Consequently, $\phi^A_{{\rm t}(Y)}$ and $v_{{\rm s}(Y)}^A$ organise into a single boundary tensor network state. Concretely, the tensor network state associated with the shape $Y = \nicefrac{(4,4,3)}{(2,1)}$ reads
\begin{equation}
    \PEPSSKEW{1}{A}{0} \, .
\end{equation}
In the following, we are interested in a special class of tensor network representations satisfying the following non-degeneracy condition:
\begin{definition}[Non-degeneracy]
    We say that a strict $\Fsq$-module category $\mc M^A$ is `non-degenerate' if $T_{1}^A$ and $T^A_2$ are faithful and dominant.      
\end{definition}
\noindent
Concretely, faithfulness means that  $T_1^A(M) \neq 0$ and $T_2^A(M) \neq 0$, for every $M \in \Irr(\mc M^A)$, whereas dominance means that every simple in $\mc M^A$ is a direct summand of $T^A_1(M_1)$, for some object $M_1$ in $\mc M^A$, and $T^A_2(M_2)$, for some object $M_2$ in $\mc M^A$. 
Therefore, $T^A_1$ is faithful and dominant if and only if, for every $M \in \Irr(\mc M^A)$, there are $M_1,M_2 \in \Irr(\mc M^A)$ such that $(T^A_1)^{M_2}_M \neq 0$ and $(T^A_1)^M_{M_1} \neq 0$. In other words, the $\Vect$-valued matrix with entries $(T^A_1)^{M_2}_{M_1}$ has no vanishing rows or columns, and similarly for $T^A_2$.

We are also interested in a special class of boundary data:

\begin{definition}[(Co)cyclicity]
    Given a strict $\Fsq$-module category $\mc M^A$, we say that boundary vectors $(\vA1,\vA2)$ are `cyclic' if 
    \begin{equation}
        \label{eq:PEPS_cyclic}
        \Span_\mathbb C \{A^i \circ v_{12}^A \, | \, i \in [d]\} = V^A_{21} \, ,
    \end{equation}
    and that boundary covectors are `cocyclic' if
    \begin{equation}
        \label{eq:PEPS_coCyclic}
        \Span_\mathbb C \{\phi_{21}^A \circ A^i  \, | \, i \in [d]\} = {V^A_{12}}^* \, .
    \end{equation}
    Furthermore, boundary data $(\vA1,\vA2,\phA1,\phA2)$ are said to have `full support' if $\vA1$ and $\phA1$ are non-zero in every block where $T^A_1$ is non-zero, and if $\vA2$ and $\phA2$ are non-zero in every block where $T^A_2$ is non-zero.
\end{definition}

\noindent
Remark for instance that, since $A^i$ preserves the labels of the two corners of the shape $Y=(1)$, the vectors $A^i \circ (v^A_{12})^{M_4}_{M_1}$ span $(T^A_2 \circ T^A_1)^{M_4}_{M_1}$ by cyclicity, for all simple objects $M_1$ and $M_4$.

Before pursuing our construction, let us briefly remark how the above cyclicity and cocyclicity requirements compare to alternative assumptions. 

\begin{remark}[MPO-injectivity]
    As alluded to in the introduction, `2-injectivity' is a stronger assumption that requires, in particular, the induced $\mathbb C$-linear map $\mathbb C^{\oplus d} \to \Nat(T_1^A \circ T_2^A,T_2^A \circ T_1^A)$ to be surjective. Together with $(T_1^A)^{M_2}_{M_1}$ and $(T_2^A)^{M_2}_{M_1}$ non-zero for every $M_1,M_2 \in \Irr(\mc M^A)$, this would guarantee that any boundary data with full support is cyclic and cocyclic. This surjectivity condition is a form of \emph{MPO-injectivity}, whereby the tensor is injective on a subspace of the bond space defined by an MPO projector. Here, the MPO simply imposes that any two adjacent legs/lines of the tensor carry the same label for the region between them.
\end{remark}

\begin{remark}[Semi-injectivity] 
    Suppose that every block $(T^A_1)_{M_1}^{M_2}$ is the tensor product of a vector space depending only on $M_1$ and a vector space depending only on $M_2$, and similarly for every block $(T^A_2)_{M_1}^{M_2}$. Therefore, every bond space $\VA1$ and $\VA2$ can be split into two halves, with one half attached to each of the adjacent regions. Collecting the halves surrounding each region, the tensor network state is obtained by placing on each plaquette a direct sum over $\Irr(\mc M)$ of a ring of maximally entangled pairs, one along each of the four aforementioned halves bounding it, and then applying at each site the map induced by the tensor.
    One verifies that the space of the four region degrees of freedom meeting at a site is dual to $\Nat(T^A_1 \circ T^A_2, T^A_2 \circ T^A_1)$ and the tensor is obtained by applying the transpose of $\mathbb C^{\oplus d} \to \Nat(T^A_1 \circ T^A_2, T^A_2 \circ T^A_1)$, which is injective by 2-injectivity.
    Therefore, in this specific case, 2-injectivity amounts to semi-injectivity in the sense of ref.~\cite{Molnar:2018hls}, whereby a tensor network state is said to be semi-injective if it can be obtained by placing a four-partite entangled state on the regions and applying an invertible $\mathbb C$-linear map at each site that act on one bond subsystem from each of the adjacent regions. Let us stress that not every semi-injective tensor network state can be obtained in this way. We speculate that this may require working over a different monoidal signature.
\end{remark}

\noindent
We illustrate our construction with a particularly simple class of tensor network states that are 2-injective but not injective: 

\begin{example}[$G$-SPTs]
    Renormalisation group fixed points of (2+1)d symmetry protected topologial phases for 0-form invertible symmetries provide examples of 2-injective tensor network representations \cite{PhysRevB.87.155114,Buerschaper:2013yfm,Williamson:2014ydz}. Let $G$ be a finite group and $\alpha$ a normalised representative of a cohomology class in $H^3(G,{\rm U}(1))$. Choose $\mc M^A$ to be the category $\Vect_G$ of $G$-graded vector spaces, regarded as a 2-vector space with $\Irr(\Vect_G) = G$, and $T^A_1 = T^A_2$ such that $(T_A^1)_{g_1}^{g_2} \cong (T_A^2)_{g_1}^{g_2} \cong \mathbb C$, for every $g_1,g_2 \in G$. This recovers the kind of GHZ entanglement structure discussed above. Taking $d = |G|^4$, we label the physical basis vectors by $i \equiv (g_1,g_2,g_3,g_4) \in G^4$ so that a copy of $\mathbb C[G]$ is assigned to every region adjacent to the tensor. Finally, we define $A^i$ by 
    \begin{equation}
        \PEPS{1}{A}{i}{g1234}{}{}{}{}{} = \delta_{i,(g_1,g_2,g_3,g_4)} \, \frac{\alpha(g_4g_2^{-1},g_2g_1^{-1},g_1)}{\alpha(g_4g_3^{-1},g_3g_1^{-1},g_1)} \, .
    \end{equation}
    Splitting the above tensors into three-valent ones, one recovers the renormalisation group fixed points discussed in ref.~\cite{PhysRevB.87.155114,Buerschaper:2013yfm,Williamson:2014ydz}. Clearly, $\mc M^A$ is 2-injective in the sense of definition \ref{def:2injectivity}. In particular, the `free' boundary data obtained by setting every component of $\vA1,\vA2,\phA1$ and $\phA2$ to $1$, are (co)cyclic.
\end{example}

\noindent
We are now ready to state the hypothesis of our fundamental theorem.

\begin{hypothesis}\label{hyp:PEPS}
    Let $\mc M^A$ and $\mc M^B$ be  two non-degenerate $\Fsq$-module categories such that $\dim_\mathbb C V_{12}^B \leq \dim_\mathbb C V^A_{12}$, $\dim_\mathbb C V_{21}^B \leq \dim_\mathbb C V^A_{21}$. Moreover, let $(\vA1,\vA2,\phA1,\phA2)$ be boundary data with full support such that $(\vA1,\vA2)$ is cyclic and $(\phA1,\phA2)$ is cocyclic. We suppose the existence of boundary data $(\vB1,\vB2,\phB1,\phB2)$ with full support such that    
    \begin{equation}    
        \label{eq:PEPS_hyp}
        \phi^A_{{\rm t}(Y)} \circ A^\fr i \circ v^A_{{\rm s}(Y)}
        = 
        \phi^B_{{\rm t}(Y)} \circ B^\fr i \circ v^B_{{\rm s}(Y)} \, ,
    \end{equation}
    for every $Y \in \Yc$ and $\fr i \in [d]^Y$.
\end{hypothesis}
\noindent
Specifically, for the shape $Y = \nicefrac{(4,4,3)}{(2,1)}$, hypothesis \ref{hyp:PEPS} stipulates that 
\begin{equation}
    \PEPSSKEW{1}{A}{1} = \hspace{-38pt}\PEPSSKEW{1}{B}{1} ,
\end{equation}
for every $i_1,\ldots,i_8 \in [d]$, where we use distinct colours for the planar regions associated with $A$ and $B$, respectively, as a visual cue. 

\begin{remark}
    Keeping all other assumptions, the condition on dimensions in hypothesis \ref{hyp:PEPS} may be replaced by the requirement that $(v_1^B,v_2^B)$ be cyclic and $(\phi^B_1,\phi^B_2)$ be cocyclic (cf. lemma~\ref{lemma:PEPS_cyclicB}).
\end{remark}

\bigskip 

\subsection{Ordinary tensor network states\label{sec:PEPS_Vec}}

Before stating and proving our fundamental theorem in the general case, let us specialise to the ordinary case. Let $\mc M^A \simeq \Vect$ and $\rho^A \colon \Fsq \to \EndC(\mc M^A)$ be a $\mathbb C$-linear strict monoidal functor. Since $\EndC(\Vect) \simeq \Vect$,  $\rho^A$ is specified by the vector spaces $\VA1$ and $\VA2$, together with the $\mathbb C$-linear maps
\begin{equation}
    A^i  \in \Hom_\mathbb C(\VA1 \otimes \VA2, \VA2 \otimes \VA1) \, , \q i \in [d] \, .
\end{equation}
Non-degeneracy of $\mc M^A$ simply states that the vector spaces $\VA1$ and $\VA2$ are non-zero.
Since $\Vect$ has a unique simple object, namely $\mathbb C$, regions carry no labels and our graphical calculus reduces to ordinary tensor network notation, i.e.,
\begin{equation}
    A^i \equiv \PEPS{0}{A}{i}{}{}{}{}{}{} \, .
\end{equation}
More generally, for every $\fr X \in \{X_1,X_2\}^*$, $V^A_{\fr X}$ is the tensor product of copies of $\VA1$ and $\VA2$, one for each letter of $\fr X$. For every shape $Y \in \Yc$ and $\fr i \in [d]^Y$, 
\begin{equation}
     A^\fr i \colon V^A_{{\rm s}(Y)} \to V^A_{{\rm t}(Y)} 
\end{equation}
is the $\mathbb C$-linear map obtained by placing at each box of $Y$ the tensor $A^i$, with $i$ the value of $\fr i$ on that box, and contracting along edges of $Y$. 

Given vectors $\vA1 \in \VA1$, $\vA2 \in \VA2$ and covectors $\phA1 \in {\VA1}^*$, $\phA2 \in {\VA2}^*$, the vector $v_\fr X^A \in V^A_{\fr X}$ reduces to the tensor product of copies of $\vA1$ and $\vA2$, one for each letter of $\fr X$, and similarly for $\phi^A_\fr X$. Cyclicity of the boundary vectors $(\vA1,\vA2)$ states  that 
\begin{equation}
    \label{eq:PEPS_Vec_Cyclic}
    \Span_\mathbb C \big\{ A^i \circ v^A_{12} \, \big| \, i \in [d]\big\} = \VA2 \otimes \VA1 \, ,
\end{equation}
while cocyclicity of the boundary covectors $(\phA1,\phA2)$ states that
\begin{equation}
    \label{eq:PEPS_Vec_coCyclic}
    \Span_\mathbb C \big\{ \phi^A_{21} \circ A^i  \, \big| \, i \in [d]\big\} = {\VA1}^* \otimes {\VA2}^* \, .
\end{equation}
Notice that cyclicity of $(\vA1,\vA2)$ and cocyclicity of $(\phA1,\phA2)$ forces $\vA1,\vA2,\phA1,\phA2 \neq 0$, which is the full support condition.
Given the above, hypothesis \ref{hyp:PEPS} boils down to the following hypothesis:
\begin{hypothesis}\label{hyp:PEPS_Vec}
    Let $A \colon \mathbb C^{\oplus d} \to \Hom_\mathbb C(\VA1 \otimes \VA2, \VA2 \otimes \VA1)$ and $B \colon \mathbb C^{\oplus d} \to \Hom_\mathbb C(\VB1 \otimes \VB2, \VB2 \otimes \VB1)$ define two (ordinary) tensor network representations on the square geometry such that $0< \dim_\mathbb C \VB1 \cdot \dim_\mathbb C \VB2 \leq \dim_\mathbb C \VA1 \cdot \dim_\mathbb C \VA2$. Moreover, let  $(\vA1,\vA2)$ be cyclic boundary vectors and $(\phA1,\phA2)$ be cocyclic boundary covectors. We suppose the existence of boundary data $(\vB1,\vB2,\phB1,\phB2)$  such that    
    \begin{equation}    
        \phi^A_{{\rm t}(Y)} \circ A^\fr i \circ v^A_{{\rm s}(Y)}
        = 
        \phi^B_{{\rm t}(Y)} \circ B^\fr i \circ v^B_{{\rm s}(Y)} \, ,
    \end{equation}
    for every $Y \in \Yc$ and $\fr i \in [d]^Y$.
\end{hypothesis}

\noindent
We quote below the fundamental theorem in a form that resembles that in which it usually appears \cite{Perez-Garcia:2010maz,Molnar:2018zys}. This theorem is a special case of our general fundamental theorem proven in the next section.

\begin{theorem}[Ordinary fundamental theorem] \label{th:PEPS_Vec_fundamental}
    Under hypothesis \ref{hyp:PEPS_Vec}, there are unique invertible maps $Q_{1} \colon \VA1 \to \VB1$ and $Q_{2} \colon \VA2 \to \VB2$ such that
    \begin{equation}
        (Q_2 \otimes Q_1) \circ A^i 
        =
        B^i \circ (Q_1 \otimes Q_2)  \, , 
    \end{equation}
    for every $i \in [d]$, and 
    \begin{equation}
        Q_1 \circ \vA1 = \vB1 \, , \q
        Q_2 \circ \vA2 = \vB2 \, , \q
        \phB1 \circ Q_1  = \phA1 \, , \q
        \phB2 \circ Q_2  = \phA2 \, .
    \end{equation}
\end{theorem}

\medskip

\subsection{Fundamental theorem\label{sec:PEPS_fundamental}}
Our goal is to establish how two tensor network representations $\mc M^A$ and $\mc M^B$ fulfilling hypothesis \ref{hyp:PEPS} are related to one another.
We begin with a simple consequence of our assumptions which, heuristically, states that upon closing off all but some of the bond degrees of freedom with the boundary data, the resulting vectors span the space associated with the remaining open bond degrees of freedom. More concretely, given a shape $Y \in \Yc$, $\fr i \in [d]^Y$ and a factorisation ${\rm t}(Y) = \fr Y \fr X \fr Y'$, we write
\begin{equation}
    \label{eq:PEPS_viX}
    \viX{A}{\fr i}{\fr X} :=
    (\phi^A_{\fr Y} \odot \id_{V^A_{\fr X}} \odot \phi^A_{\fr Y'}) \circ A^\fr i \circ v^A_{{\rm s}(Y)} \in V_\fr X^A \, ,
\end{equation}
where $\phi^A_{\fr Y} \odot \id_{V^A_{\fr X}} \odot \phi^A_{\fr Y'}$ denotes the map whose transpose is $\phi \mapsto \phi^A_{\fr Y} \odot \phi \odot \phi^A_{\fr Y'}$. Dually, given a factorisation ${\rm s}(Y) = \fr Z \fr X' \fr Z'$,
\begin{equation}
    \label{eq:PEPS_phiX}
    \phiX{A}{\fr i}{\fr X'} := 
    \phi^A_{{\rm t}(Y)} \circ A^\fr i \circ (v_\fr Z^A \odot \id_{V^A_{\fr X'}} \odot v_{\fr Z'}^A) \, ,
\end{equation}
where $v_\fr Z^A \odot \id_{V^A_{\fr X'}} \odot v_{\fr Z'}^A$ denotes the map $v \mapsto v_\fr Z^A \odot v \odot v_{\fr Z'}^A$. We define $\viX{B}{\fr i}{\fr X}$ and $\phiX{B}{\fr i}{\fr X'}$ similarly. 
We leave the factorisation implicit in the notation, as it will always be clear from the context. Formally, the above statement is the content of the following lemma:

\begin{lemma} \label{lemma:PEPS_Span}
    Let $\mc M^A$ be non-degenerate. Let $(\vA1,\vA2,\phA1,\phA2)$ be boundary data with full support such that $(\vA1,\vA2)$ is cyclic and $(\phA1,\phA2)$ is cocyclic, and let $Y$ be a shape. 
    For every factorisation ${\rm t}(Y) = \fr Y \fr X \fr Y'$,
    \begin{align}
        \Span_\mathbb C \big\{ \viX{A}{\fr i}{\fr X} \, \big| \, \fr i \in [d]^Y\big\} = V^A_{\fr X} \, .
    \end{align}
    For every factorisation ${\rm s}(Y) = \fr Z \fr X' \fr Z'$,
    \begin{equation}
        \Span_\mathbb C \big\{ \phiX{A}{\fr i}{\fr X'} \, \big| \, \fr i \in [d]^Y\big\} = {V^A_{\fr X'}}^* \, ,
    \end{equation}

\end{lemma}
\noindent
\begin{proof}
    Let us prove the first assertion.  
    First, observe that 
    \begin{equation}
        V^A_{{\rm t}(Y)} 
        \cong
        \bigoplus_{M_1,\ldots,M_4} 
        (T^A_\fr Y)_{M_3}^{M_4} \otimes 
        (T^A_{\fr X})_{M_2}^{M_3} \otimes 
        (T^A_{\fr Y'})_{M_1}^{M_2} \, . 
    \end{equation}
    Fix $M_2,M_3 \in \Irr(\mc M^A)$. Since $\mc M^A$ is non-degenerate, there are $M_1,M_4 \in \Irr(\mc M^A)$ such that the blocks $(T^A_\fr Y)_{M_3}^{M_4}$ and $(T^A_{\fr Y'})_{M_1}^{M_2}$ are non-zero. 
    Moreover, since the boundary data have full support, the components $(\phi^A_\fr Y)^{M_4}_{M_3}$ and $(\phi^A_{\fr Y'})^{M_2}_{M_1}$ are non-zero.  
    As this holds for all $M_2,M_3$, the map
    \begin{equation}
        \phi^A_{\fr Y} \odot \id_{V^A_{\fr X}} \odot \phi^A_{\fr Y'} \colon V^A_{{\rm t}(Y)} \to V^A_{\fr X}
    \end{equation}
    is surjective. Therefore, it is enough to prove that
    \begin{align}
        \Span_\mathbb C \big\{  A^\fr i \circ v^A_{{\rm s}(Y)} \, \big| \, \fr i \in [d]^Y\big\} = V^A_{{\rm t}(Y)} \, .
    \end{align}
    Build $Y$ box by box, each box coming after the boxes directly below it and to its right. Correspondingly, the first tensor in the network always acts on $v_{12}^A$, so by cyclicity the resulting vectors span $V^A_{21} \ni \vA2 \odot \vA1$. 
    Thus, each subsequent tensor again acts on $v_{12}^A = v^A_1 \odot v^A_2$, amongst other vectors. 
    Applying eq.~\eqref{eq:PEPS_cyclic} at every step eventually yields all of $V^A_{{\rm t}(Y)}$. The second assertion follows in the same way from cocyclicity. 
\end{proof}

\medskip
\noindent 
For instance, lemma \ref{lemma:PEPS_Span} implies that the $\mathbb C$-linear maps defined via
\begin{align}
    e_i \mapsto \! \PEPS{1}{A}{i}{}{}{v}{v}{\phi}{} \equiv \viX{A}{i}{X_1}
    \;\; \text{and} \q
    e_i \mapsto \, \PEPS{1}{A}{i}{}{}{v}{v}{}{\phi} \equiv \viX{A}{i}{X_2} \, ,
\end{align}
respectively, are surjective. 

\noindent

\begin{lemma} \label{lemma:PEPS_cyclicB}
    Under hypothesis \ref{hyp:PEPS}$, (\vB1,\vB2)$ is cyclic and $(\phB1,\phB2)$ is cocyclic. 
\end{lemma}
\noindent
\begin{proof}
    Our strategy is the same as in the one-dimensional case. 
    Consider the shape $Y := \nicefrac{(2,2)}{(1)}$, for which ${\rm s}(Y) = X_1X_1X_2X_2$ and ${\rm t}(Y) = X_2 X_1 X_2 X_1$. To every $\fr i \in [d]^Y$, one associates the complex number $\phi^A_{{\rm t}(Y)} \circ A^\fr i \circ v^A_{{\rm s}(Y)}$. This extends to a linear map
    \begin{equation}
        \label{eq:PEPS_221}
        \big( e_\fr i \mapsto \phi^A_{{\rm t}(Y)} \circ A^\fr i \circ v^A_{{\rm s}(Y)} \big) \equiv 
        \PEPSC{1}{}{}{}{}{}{\phi}{} 
    \end{equation}
    which we regard as a bilinear form on $\mathbb C^{\oplus d} \times  (\mathbb C^{\oplus d})^{\otimes 2}$, where the first factor is associated with the bottom-right tensor.
    This form pairs $A^{i_1} \circ v_{12}^A$, which span $V^A_{21}$ by cyclicity, with the covectors 
    \begin{equation}
        \label{eq:PEPS_221_half}
        \big(\phi_{21}^A \circ A^{i_2} \circ (v_1^A \odot \id_{\VA2}) \big) \odot \big( \phi^A_{21} \circ A^{i_3} \circ (\id_{\VA1} \odot \vA2)\big) \equiv \phiX{A}{i_2}{X_2} \odot \phiX{A}{i_3}{X_1} \, .
    \end{equation}
    Indeed, the bottom-left and top-right tensors share a single region, along whose label the two factors in eq.~\eqref{eq:PEPS_221_half} are contracted; these share three regions with the bottom-right tensor, whose labels are summed over in the pairing of ${V_{21}^A}^*$ with $V_{21}^A$.
    By lemma \ref{lemma:PEPS_Span}, the factors in eq.~\eqref{eq:PEPS_221_half} span ${\VA2}^*$ and ${\VA1}^*$, respectively, so that their $\odot$-products span ${V_{21}^A}^*$. 
    Therefore, the rank of the map \eqref{eq:PEPS_221} equals $\dim_\mathbb C V^A_{21}$. By hypothesis \ref{hyp:PEPS}, the form obtained by replacing $A$ by $B$ possesses the same rank. But, the same factorisation shows that it factors through $\Span_\mathbb C \{B^i \circ v_{12}^B \, | \, i \in [d]\}$, so that
    \begin{equation}
        \dim_\mathbb C V^A_{21} \leq \dim_\mathbb C \Span_\mathbb C \big\{B^i \circ v_{12}^B \, \big| \, i \in [d]\big\} \leq \dim_\mathbb C V_{21}^B  \, .
    \end{equation}
    Since $\dim_\mathbb C V^B_{21} \leq \dim_\mathbb C V^A_{21}$ by hypothesis \ref{hyp:PEPS}, this forces $\Span_\mathbb C \big\{B^i \circ v_{12}^B \, \big| \, i \in [d]\big\} = V^B_{21}$, i.e. $(\vB1,\vB2)$ cyclic. The same argument with the shape $Y := (2,1)$ yields $(\phB1,\phB2)$ cocyclic.
\end{proof}

\medskip \noindent
Recall that the source of a shape begins with $X_1$ and ends with $X_2$, whereas its target begins with $X_2$ and ends with $X_1$. For any non-empty word $\fr X$, consider two shapes $Y,Y' \in \Yc$ and factorisations ${\rm t}(Y) = \fr Y \fr X \fr Y'$ and ${\rm s}(Y') = \fr Z \fr X \fr Z'$. Provided that $\fr Y$ or $\fr Z$ is empty, and $\fr Y'$ or $\fr Z'$ is empty, identifying the factor $\fr X$ of ${\rm t}(Y)$ with that of ${\rm s}(Y')$ produces a shape with source $\fr Z {\rm s}(Y) \fr Z'$ and target $\fr Y {\rm t}(Y') \fr Y'$. Equivalently, the morphisms 
\begin{equation}    
    \label{eq:PEPS_gluing_mor}
    \id_{\fr Z} \otimes e_{\fr i} \otimes \id_{\fr Z'}
    \q \text{and} \q 
    \id_{\fr Y} \otimes e_{\fr i'} \otimes \id_{\fr Y'}
\end{equation}    
are composable in $\Fsq$, for every $\fr i \in [d]^Y$ and $\fr i' \in [d]^{Y'}$. In that case, we say that $Y$ and $Y'$ are `glued' along $\fr X$, denote the resulting shape by $Y \sqcup Y'$ and write $\fr i \sqcup \fr i' \in [d]^{Y \sqcup Y'}$ for the function such that $e_{\fr i \sqcup \fr i'}$ is the composite of the morphisms \eqref{eq:PEPS_gluing_mor} and hence (cf. eq.~\eqref{eq:PEPS_idAid})
\begin{equation}
    A^{\fr i \sqcup \fr i'} = (\id_{V^A_\fr Y} \odot A^{\fr i'} \odot \id_{V^A_{\fr Y'}}) \circ (\id_{V^A_\fr Z} \odot A^{\fr i} \odot \id_{V^A_{\fr Z'}}) \, .
\end{equation}
Note that it is not possible for $\fr Y$ and $\fr Z$ to be both empty. Therefore, if $Y$ and $Y'$ are glued along $\fr X$, exactly one of $\fr Y,\fr Z$ is empty;  $\fr Y$ is empty if $\fr X$ begins with $X_2$, and $\fr Z$ is empty if $\fr X$ begins with $X_1$. Similarly, exactly one of $\fr Y',\fr Z'$ is empty, namely $\fr Y'$ if $\fr X$ ends with $X_1$ and $\fr Z'$ if $\fr X$ ends with $X_2$.

\begin{definition}[Exposed words] \label{def:PEPS_exposed}
    Let $\fr X$ be a non-empty word in $\{X_1,X_2\}^*$. Given a shape $Y$ with ${\rm t}(Y) = \fr Y \fr X \fr Y'$, we say that $\fr X$ is `exposed' in ${\rm t}(Y)$ if there is a shape $Y'$ with ${\rm s}(Y') = \fr Z \fr X \fr Z'$ such that $Y$ and $Y'$ can be glued along $\fr X$. Dually, given a shape $Y'$ with ${\rm s}(Y') = \fr Z \fr X \fr Z'$, we say that $\fr X$ is exposed in ${\rm s}(Y')$ if there is a shape $Y$ with ${\rm t}(Y) = \fr Y \fr X \fr Y'$ such that $Y$ and $Y'$ can be glued along $\fr X$. 
\end{definition}

\noindent
The point of the above definition is that if $\fr X$ is exposed in both ${\rm t}(Y)$ and ${\rm s}(Y')$, then $Y$ and $Y'$ can be glued along $\fr X$. Indeed, by the above characterisations, if $\fr X$ begins with $X_2$ then $\fr Y$ is empty since $\fr X$ is exposed in ${\rm t}(Y)$, whereas if it begins with $X_1$, then $\fr Z$ is empty since $\fr X$ is exposed in ${\rm s}(Y')$; and similarly for $\fr Y'$ and $\fr Z'$.  Every non-empty word $\fr X$ is exposed in some target and in some source. Indeed, adding $X_2$ at the front of $\fr X$ if it begins with $X_1$, and $X_1$ at the back if it ends with $X_2$, yields the target ${\rm t}(Y)$ of some shape $Y$. Similarly, adding $X_1$ at the front of $\fr X$ if it begins with $X_2$, and $X_2$ at the back if it ends with $X_1$, yields the source ${\rm s}(Y')$ of some shape $Y'$. Then, $Y$ and $Y'$ can be glued along $\fr X$.
In particular, ${\rm s}(Y)$ and ${\rm t}(Y)$ are exposed in themselves, for every shape $Y$.

We are now ready to prove the statement that constitutes the nuts and bolts of our strategy. 
\begin{proposition} \label{prop:PEPS_Q}
    Let $\fr X$ be a non-empty word in $\{X_1,X_2\}^*$. Under hypothesis \ref{hyp:PEPS}, there is a unique invertible $\mathbb C$-linear map $Q_\fr X \colon V^A_{\fr X} \to V^B_{\fr X}$ such that (cf. notation of eqs.~\eqref{eq:PEPS_viX} and \eqref{eq:PEPS_phiX})
    \begin{equation}
        \label{eq:PEPS_Q_action}
        Q_\fr X \circ \viX{A}{\fr i}{\fr X}
        =
        \viX{B}{\fr i}{\fr X} 
    \end{equation}
    holds for every shape $Y$ with ${\rm t}(Y) = \fr Y \fr X \fr Y'$ in which $\fr X$ is exposed, and every $\fr i \in [d]^Y$. Moreover, it satisfies
    \begin{equation}
        \label{eq:PEPS_Q_action_bis}
        \phiX{B}{\fr i'}{\fr X} \circ Q_\fr X = \phiX{A}{\fr i'}{\fr X}
    \end{equation}
    for every shape $Y'$ with ${\rm s}(Y') = \fr Z \fr X \fr Z'$ in which $\fr X$ is exposed, and $\fr i' \in [d]^{Y'}$.
\end{proposition}
\noindent
\begin{proof}
    Let $Y'$ be a shape with ${\rm s}(Y') = \fr Z \fr X \fr Z'$ in which $\fr X$ is exposed, which always exists. For every shape $Y$ with ${\rm t}(Y)= \fr Y \fr X \fr Y'$ in which $\fr X$ is exposed, lemma \ref{lemma:PEPS_Span} yields $\Span_\mathbb C \{\viX{A}{\fr i}{\fr X} \, | \, \fr i \in [d]^Y\} = V^A_{\fr X}$. Therefore, the assignment $\viX{A}{\fr i}{\fr X} \mapsto \viX{B}{\fr i}{\fr X}$ defines $Q_\fr X$ everywhere, provided it is well defined.  
    Since $\fr X$ is exposed in both ${\rm t}(Y)$ and ${\rm s}(Y')$, the shapes $Y$ and $Y'$ can be glued along $\fr X$.
    The tensors associated with $Y$ and $Y'$ in $Y \sqcup Y'$ share exactly the regions adjacent to the edges carrying the letters of $\fr X$, whose labels are summed over in the pairing of ${V^A_\fr X}^*$ with $V^A_\fr X$, and similarly for $B$. Hence,
    \begin{align}
        \phiX{A}{\fr i'}{\fr X} \! \circ \viX{A}{\fr i}{\fr X} &= \phi^A_{{\rm t}(Y \sqcup Y')} \circ A^{\fr i \sqcup \fr i'} \circ v^A_{{\rm s}(Y 
        \sqcup Y')} \, ,
        \\
        \phiX{B}{\fr i'}{\fr X} \! \circ \viX{B}{\fr i}{\fr X} &= \phi^B_{{\rm t}(Y \sqcup Y')} \circ B^{\fr i \sqcup \fr i'} \circ v^B_{{\rm s}(Y 
        \sqcup Y')} \, .
    \end{align}
    By hypothesis \ref{hyp:PEPS}, it follows that
    \begin{equation}
        \phiX{A}{\fr i'}{\fr X} \! \circ \viX{A}{\fr i}{\fr X} = \phiX{B}{\fr i'}{\fr X} \! \circ \viX{B}{\fr i}{\fr X} \, ,
    \end{equation}
    for every such $Y$, $\fr i \in [d]^Y$ and $\fr i' \in [d]^{Y'}$. Now, suppose that $\sum_{\fr i}c_\fr i \, \viX{A}{\fr i}{\fr X} = 0$ for some coefficients $c_\fr i \in \mathbb C$, where the finite sum runs over functions $\fr i$ of possibly different such shapes $Y$ whose targets $\fr Y \fr X \fr Y'$ may also differ provided that $\fr X$ is exposed in them. It follows from the above equality that
    \begin{equation}
        \phiX{B}{\fr i'}{\fr X} \circ \bigg( \sum_{\fr i}c_\fr i \, \viX{B}{\fr i}{\fr X}\bigg)
        =
        \phiX{A}{\fr i'}{\fr X}\circ \bigg( \sum_{\fr i}c_\fr i \, \viX{A}{\fr i}{\fr X}\bigg) = 0 \, ,
    \end{equation}
    for every $\fr i' \in [d]^{Y'}$.
    By lemmas \ref{lemma:PEPS_cyclicB} and \ref{lemma:PEPS_Span}, we have $\Span_\mathbb C \{\phiX{B}{\fr i'}{\fr X} \, | \, \fr i' \in [d]^{Y'}\} = {V^B_{\fr X}}^*$. Hence $\sum_{\fr i}c_\fr i \, \viX{B}{\fr i}{\fr X} = 0$, which makes $Q_\fr X$ well defined, independently of the shape $Y$. Exchanging the roles of $A$ and $B$ produces a well-defined map $V^B_{\fr X} \to V^A_{\fr X}$, $\viX{B}{\fr i}{\fr X} \mapsto \viX{A}{\fr i}{\fr X}$, which is the inverse of $Q_\fr X$.
    Finally, 
    \begin{equation}
        \phiX{B}{\fr i'}{\fr X} \circ Q_\fr X \circ \viX{A}{\fr i}{\fr X} = \phiX{A}{\fr i'}{\fr X}\circ \viX{A}{\fr i}{\fr X} \, ,
    \end{equation}
    for every $\fr i \in [d]^Y$. By lemma \ref{lemma:PEPS_Span},  $\Span_\mathbb C \{\viX{A}{\fr i}{\fr X} \, | \, \fr i \in [d]^{Y}\} = V^A_\fr X$. Hence, $\phiX{B}{\fr i'}{\fr X} \circ Q_\fr X = \phiX{A}{\fr i'}{\fr X}$, for every $\fr i' \in [d]^{Y'}$. Since $Q_\fr X$ is solely defined by eq.~\eqref{eq:PEPS_Q_action}, it does not depend  on the choice of $Y'$, so that eq.~\eqref{eq:PEPS_Q_action_bis} holds for every $Y'$ as desired. 
\end{proof}

\medskip \noindent
Let us immediately apply proposition \ref{prop:PEPS_Q} to several configurations. 
Consider the gluing of the box $Y'=(1)$ on top of the box $Y=(1)$. It shows that factor $X_1$ is exposed in both $X_2X_1 = {\rm t}((1))$ and $X_1X_2 = {\rm s}((1))$. Proposition \ref{prop:PEPS_Q} thus yields an invertible map $Q_1 \equiv Q_{X_1} \colon \VA1 \xrightarrow{\sim} \VB1$ such that
\begin{align}
    Q_1 \circ (\phA2 \odot {\rm id}_{\VA1}) \circ A^i \circ v^A_{12}
    &=
    (\phB2 \odot {\rm id}_{\VB1}) \circ B^i \circ v^B_{12} \, ,
    \\
    \phi^B_{21} \circ B^i \circ (\id_{\VB1} \odot \vB2) \circ Q_1 &= \phi^A_{21} \circ A^i \circ (\id_{\VA1} \odot \vA2) \, ,
\end{align}
for every $i \in [d]$. 
In tensor network notation, we depict these equations as follows: 
\begin{equation}
    \label{eq:PEPS_Q1}
    \PEPS{1}{A}{}{}{}{v}{v}{\phi}{Q} = \PEPS{1}{B}{}{}{}{v}{v}{\phi}{} 
    , \q
    \PEPS{1}{B}{}{}{}{Q}{v}{\phi}{\phi} = \PEPS{1}{A}{}{}{}{}{v}{\phi}{\phi} .
\end{equation}
Let us spell out once the meaning of such pictures. For instance, for every $M_0,M_6 \in \Irr(\mc M^B)$, we have
\begin{equation}
\begin{split}
    \PEPS{1}{A}{}{}{M60}{v}{v}{\phi}{Q} 
    &= 
    \sum_{M_1,\ldots,M_4 \in \Irr(\mc M^A)} \!\!\!\!\! 
    \PEPS{1}{A}{}{M1234}{M60}{v}{v}{\phi}{Q} \\ 
    &= 
    \sum_{M_5,M_7 \in \Irr(\mc M^B)} \PEPS{1}{B}{}{M0567}{}{v}{v}{\phi}{}  = \PEPS{1}{B}{}{M0-6-}{}{v}{v}{\phi}{} .
\end{split}
\end{equation}
Similarly, proposition \ref{prop:PEPS_Q} yields an invertible map $Q_2 \equiv Q_{X_2} \colon \VA2 \xrightarrow{\sim} \VB2$ such that
\begin{align}
    Q_2 \circ ({\rm id}_{\VA2} \odot \phA1) \circ A^i \circ v^A_{12}
    &=
    ({\rm id}_{\VB2} \odot \phB1) \circ B^i \circ v^B_{12} \, ,
    \\
    \phi^B_{21} \circ B^i \circ (\vB1 \odot \id_{\VB2}) \circ Q_2 &= \phi^A_{21} \circ A^i \circ (\vA1 \odot \id_{\VA2}) \, ,
\end{align}
for every $i \in [d]$, which in tensor network notation reads
\begin{equation}
    \label{eq:PEPS_Q2}
    \PEPS{1}{A}{i}{}{}{v}{v}{Q}{\phi} = \PEPS{1}{B}{i}{}{}{v}{v}{}{\phi} 
    , \q
    \PEPS{1}{B}{i}{}{}{v}{Q}{\phi}{\phi} = \PEPS{1}{A}{i}{}{}{v}{}{\phi}{\phi} \, .
\end{equation}

\begin{lemma}   \label{lemma:PEPS_Qv}
    Under hypothesis \ref{hyp:PEPS}, the aforementioned invertible maps $Q_1$ and $Q_2$ satisfy
    \begin{equation}
        Q_1 \circ \vA1 = \vB1 \, , \q
        Q_2 \circ \vA2 = \vB2 \, , \q
        \phB1 \circ Q_1  = \phA1 \, , \q
        \phB2 \circ Q_2  = \phA2 \, .
    \end{equation}
\end{lemma}
\noindent
\begin{proof}
    By eq.~\eqref{eq:PEPS_Q1} and under hypothesis \ref{hyp:PEPS},
    \begin{equation}
    \begin{split}
        \phi^B_{21} \circ B^i \circ (\id_{\VB1} \odot \vB2) \circ Q_1 \circ \vA1 &= \phi_{21}^A \circ A^i \circ v^A_{12} 
        \\
        &= \phi_{21}^B \circ B^i \circ v^B_{12}  = \phi^B_{21} \circ B^i \circ (\id_{\VB1} \odot \vB2) \circ \vB1 \, .            
    \end{split}
    \end{equation}
    By lemmas \ref{lemma:PEPS_cyclicB} and \ref{lemma:PEPS_Span}, we have $\Span_\mathbb C \{ \phi^B_{21} \circ B^i \circ (\id_{\VB1} \odot \vB2)\, | \, i \in [d]\} = {\VB1}^*$, which yields $Q_1 \circ \vA1 = \vB1$. The other statements can be obtained in the same way.
\end{proof}

\medskip \noindent
Next, gluing a box $Y=(1)$ into the notch of the shape $Y'=(2,1)$ shows that $X_2X_1$ is exposed in both $X_2X_1 = {\rm t}((1))$ and $X_1X_2X_1X_2={\rm s}((2,1))$.\footnote{\label{foot:PEPS_X2X1}Note that the word $X_2 X_1$ is also a factor of the target of the shapes $(2,2)$ and $(2,1)$, but it is not exposed inside them. Since both $\fr Y = X_2$ and $\fr Y'=X_1$ are non-empty, a shape $Y'$ as in definition \ref{def:PEPS_exposed} would need to satisfy ${\rm s}(Y')=X_2 X_1$, which is impossible. In other words, these shapes cannot be used to define the map $Q_{21}$.} Similarly, gluing a box $Y'=(1)$ into the notch of the shape $Y=\nicefrac{(2,2)}{(1)}$ shows that $X_1X_2$ is exposed in both $X_2X_1X_2X_1 = {\rm t}(\nicefrac{(2,2)}{(1)})$ and $X_1X_2 = {\rm s}((1))$. 
Proposition \ref{prop:PEPS_Q} thus yields invertible maps $Q_{21} \equiv Q_{X_2X_1} \colon V^A_{21} \xrightarrow{\sim} V^B_{21}$ and $Q_{12} \equiv Q_{X_1X_2} \colon V^A_{12} \xrightarrow{\sim} V^B_{12}$, which satisfy in particular
\begin{equation}
    \label{eq:PEPS_Q21}
    \PEPSC{1}{i_1}{}{}{}{Q21}{}{1}
     = 
    \PEPS{1}{B}{i_1}{}{}{v}{v}{}{}
    , \q 
   \PEPSCB{1}{i_2}{}{}{}{Q21}{}{1} = \PEPS{1}{A}{i_2}{}{}{}{}{\phi}{\phi} \, ,
\end{equation}
for every $i_1,i_2 \in [d]$. The next two lemmas are about expressing $Q_{21}$ and $Q_{12}$ in terms of $Q_1$ and $Q_2$.

\begin{lemma}\label{lemma:PEPS_Q1Q2}
    Under hypothesis \ref{hyp:PEPS}, the invertible maps $Q_{12}$ and $Q_{21}$ satisfy\footnote{Recall from sec.~\ref{sec:PEPS_states} that $\iota^A \colon V^A_{\fr X \fr X'} \to V^A_{\fr X} \otimes V^A_{\fr X'}$ is the inclusion map and $\pi^A \colon V^A_{\fr X} \otimes V^A_{\fr X'} \to V^A_{\fr X \fr X'}$ the projetion $v \otimes v' \mapsto v \odot v'$.}    
    \begin{align}
        \label{eq:PEPS_Q12_pi}
        \pi^B \circ (Q_1 \otimes Q_2)  
        &= Q_{12} \circ \pi^A \colon V^A_1 \otimes V^A_2 \to V^B_{12} \, ,
        \\
        \label{eq:PEPS_Q21_iota}
        (Q_2 \otimes Q_1) \circ \iota^A 
        &= \iota^B \circ Q_{21} \colon V^A_{21} \to V^B_2 \otimes V^B_1 \, .
    \end{align}
\end{lemma}

\noindent
\begin{proof}
    Let us focus on eq.~\eqref{eq:PEPS_Q21_iota}. Consider the shape $Y = \nicefrac{(2,2)}{(1)}$. By eq.~\eqref{eq:PEPS_Q1}, eq.~\eqref{eq:PEPS_Q2}, eq.~\eqref{eq:PEPS_Q21} and under hypothesis \ref{hyp:PEPS}, we have\footnote{Notice that, in the final result, the tensors associated with $B$ do not share any region.}
    \begin{equation}
    \begin{split}
        \PEPSC{1}{i_1}{i_2}{i_3}{}{Q21}{\phi}{1} 
        &=
        \PEPSC[B]{1}{i_1}{i_2}{i_3}{}{}{\phi}{1}
        \\
        &=
        \PEPSC[A]{1}{i_1}{i_2}{i_3}{}{}{\phi}{1}
        = \PEPSC{1}{i_1}{i_2}{i_3}{}{Q2Q1}{\phi}{1} \, ,
    \end{split}
    \end{equation}
    for every $i_1,i_2,i_3 \in [d]$.
    By lemmas \ref{lemma:PEPS_Span} and \ref{lemma:PEPS_cyclicB}, the covectors associated with the tensors labelled by $i_2$ and $i_3$ span ${\VB2}^*$ and ${\VB1}^*$, respectively. By cyclicity, the vectors $A^{i_1} \circ v_{12}^A$ span $V_{21}^A$. 
    Therefore,
    \begin{equation}
        \label{eq:PEPS_Q21_covectors}
        (\phi' \odot \phi) \circ Q_{21}  = (\phi' \circ Q_2) \odot (\phi \circ Q_1) \, ,
    \end{equation}
    for every $\phi \in {\VB1}^*$ and $\phi' \in {\VB2}^*$. In the same way, the shape $(2,1)$ yields
    \begin{equation}
        \label{eq:PEPS_Q12_vectors}
        Q_{12} \circ (v \odot v') = (Q_1 \circ v) \odot (Q_2 \circ v') \, ,
    \end{equation}
    for every $v \in \VA1$ and $v' \in 
    \VA2$.
    Graphically,\footnote{Notice how some of the regions are split in two; each half carries its own label.}
    \begin{equation}
        \QTTLine{Q12}{}{v}{v'}{}{}
         \!\! = \QTTLine{Q1Q2a}{}{v}{v'}{}{} \, , \q
         \QTTLine{Q21}{}{}{}{\phi'}{\phi}
         = \!\! \QTTLine{Q2Q1b}{}{}{}{\phi'}{\phi} \, .
    \end{equation}
    Regard $V_{21}^A$ and $V_{21}^B$ as subspaces of $\VA2 \otimes \VA1$ and $\VB2 \otimes \VB1$ by eq.~\eqref{eq:PEPS_VAXX}, respectively. Since a $\otimes$-product of covectors restricts to their $\odot$-product on both $V_{21}^A$ and $V_{21}^B$, eq.~\eqref{eq:PEPS_Q21_covectors} implies that 
    \begin{equation}
    \begin{split}
        (\phi' \otimes \phi) \circ Q_{21} \circ (v' \odot v) &= 
        \big((\phi' \circ Q_2) \odot (\phi \circ Q_1)\big) \circ (v' \odot v)
        \\
        &=(\phi' \otimes \phi) \circ (Q_2 \otimes Q_1) \circ (v' \odot v) \, ,
    \end{split}
    \end{equation}
    for every $v \in \VA1$ and $v' \in \VA2$. Since the covectors $(\phi' \otimes \phi)$ span $(\VB2 \otimes \VB1)^*$, we have $Q_{21} \circ (v' \odot v) = (Q_2 \otimes Q_1) \circ (v' \odot v)$, for every $v \in \VA1$ and $v' \in \VA2$. 
    Given that the vectors $v' \odot v$ span $V_{21}^A$, one concludes that $Q_{21}$ is the restriction of $Q_2 \otimes Q_1$ to $V_{21}^A$, which is eq.~\eqref{eq:PEPS_Q21_iota}. Similarly, since $\pi^A \circ (v \otimes v') = v \odot v'$ and $\pi^B \circ ((Q_1 \circ v) \otimes (Q_2 \circ v')) = (Q_1 \circ v) \odot (Q_2 \circ v')$, eq.~\eqref{eq:PEPS_Q12_vectors} implies that
    \begin{equation}
        Q_{12} \circ \pi^A \circ (v \otimes v') = \pi^B \circ (Q_1 \otimes Q_2) \circ (v \otimes v') \, ,
    \end{equation}
    for every $v \in \VA1$ and $v' \in \VA2$. Since the vectors $v \otimes v'$ span $\VA1 \otimes \VA2$, one concludes eq.~\eqref{eq:PEPS_Q12_pi}.
\end{proof}

\medskip
\begin{lemma}
    \label{lemma:PEPS_bijections}
    Under hypothesis \ref{hyp:PEPS}, there is a bijection $\sigma \colon \Irr(\mc M^A) \to \Irr(\mc M^B)$ such that $Q_1$ maps $(T^A_1)_{M_1}^{M_2}$ to $(T^B_1)_{\sigma(M_1)}^{\sigma(M_2)}$ and $Q_2$ maps $ (T^A_2)_{M_1}^{M_2}$ to $ (T^B_2)_{\sigma(M_1)}^{\sigma(M_2)}$.
\end{lemma}

\noindent
\begin{proof}
    By lemma \ref{lemma:PEPS_Q1Q2}, $Q_2 \otimes Q_1$ maps $V^A_{21}$ to $V^B_{21}$. Graphically,
    \begin{equation}
        \label{eq:PEPS_Q21Q2Q1}
        \QTTLine{Q21}{}{}{}{}{}
        =
        \QTTLine{Q2Q1b}{}{}{}{}{}
        = \!\!
        \bigoplus_{M_5,M_6 \in \Irr(\mc M^B)}
        \QTTLine{Q2Q1b}{M------65}{}{}{}{}
        = \bigoplus_{\substack{M_5 \in \Irr(\mc M^B) \\ M_2 \in \Irr(\mc M^A)}} \QTTLine{Q2Q1b}{M-2----55}{}{}{}{} \equiv 
        \QTTLine{Q2Q1}{}{}{}{}{} \, ,
    \end{equation}
    where the third equality holds since $(Q_2 \otimes Q_1)(V_{21}^A) = V_{21}^B$ forces the terms with $M_5 \neq M_6 \in \Irr(\mc M^B)$ to vanish. Now, fixing $M_2 \in \Irr(\mc M^A)$,
    each term in the sum is the tensor product of the blocks 
    \begin{equation}
        \bigoplus_{M_1} \,(T^A_1)_{M_1}^{M_2} \to \bigoplus_{M_4} \,(T^B_1)_{M_4}^{M_5} 
        \q \text{and} \q 
        \bigoplus_{M_3} \,(T^A_2)_{M_2}^{M_3} \to \bigoplus_{M_7} \,(T^B_2)_{M_6}^{M_7}
    \end{equation}
    of $Q_1$ and $Q_2$, respectively. 
    Therefore, since the terms with $M_5 \neq M_6$ vanish, the block of $Q_1$ labelled by $M_5$ and the block of $Q_2$ labelled by $M_6$ can only be simultaneously non-zero if $M_5 = M_6$.
    By non-degeneracy of $\mc M^A$ and injectivity of $Q_1,Q_2$, each block must be non-zero for at least one label.
    All in all, there is a unique label $\sigma(M_2) \in \Irr(\mc M^B)$ for which the block of $Q_1$ is non-zero if and only if $M_5 = \sigma(M_2)$, and the block of $Q_2$ is non-zero if and only if $M_6 = \sigma(M_2)$.
    Hence,
    \begin{equation}
        \QSQ{Q1}{}{}{}{}
        = \!\!\! \bigoplus_{M \in \Irr(\mc M^A)}
        \QSQ{Q1}{}{M}{}{\sigma(M)} 
        \, , \q
        \QSQ{Q2}{}{}{}{}
        = \!\!\! \bigoplus_{M \in \Irr(\mc M^A)}
        \QSQ{Q2}{M}{}{\sigma(M)}{} \, .
    \end{equation}
    The same reasoning applied to $Q_{21}^{-1}$, $Q_2^{-1}$ and $Q_1^{-1}$ produces the inverse of the map  $\sigma \colon \Irr(\mc M^A) \to \Irr(\mc M^B)$.\footnote{Note that the existence of a bijection $\sigma \colon \Irr(\mc M^A) \to \Irr(\mc M^B)$ already implies, by semisimplicity, that the 2-vector spaces $\mc M^A$ and $\mc M^B$ are equivalent.}
    Transposing eq.~\eqref{eq:PEPS_Q12_pi} shows that $Q_{12}^*$ is the restriction of $Q_1^* \otimes Q_2^*$ to $(V^B_{12})^*$, so that the above reasoning applies to the transposed maps. 
    This yields another bijection $\sigma' \colon \Irr(\mc M^A) \to \Irr(\mc M^B)$ such that
    \begin{equation}
        \QSQ{Q1}{}{}{}{}
        = \!\!\! \bigoplus_{M \in \Irr(\mc M^A)}
        \QSQ{Q1}{M}{}{\sigma'\! (M)}{} 
        \, , \q
        \QSQ{Q2}{}{}{}{}
        = \!\!\! \bigoplus_{M \in \Irr(\mc M^A)}
        \QSQ{Q2}{}{M}{}{\sigma'\! (M)} \, .
    \end{equation}
    The combination of these mappings yields
    \begin{align}
        \label{eq:PEPS_Q1_blocks}
        Q_1 &= 
        \!\!\! \bigoplus_{M_1,M_2 \in \Irr(\mc M^A)} \!\!\!
        (Q_1)_{M_1}^{M_2} 
        \q \text{with} \q
        (Q_1)_{M_1}^{M_2} \colon (T^A_1)_{M_1}^{M_2} \to (T^B_1)_{\sigma'(M_1)}^{\sigma(M_2)} \, ,
        \\
        \label{eq:PEPS_Q2_blocks}
        Q_2 &= 
        \!\!\! \bigoplus_{M_1,M_2 \in \Irr(\mc M^A)} \!\!\!
        (Q_2)_{M_1}^{M_2} 
        \q \text{with} \q
        (Q_2)_{M_1}^{M_2} \colon (T^A_2)_{M_1}^{M_2} \to (T^B_2)_{\sigma(M_1)}^{\sigma'(M_2)} \, .
    \end{align}
    Defining $Q_2 \odot Q_1 \colon V_{21}^A \to V_{21}^B$ and $Q_1 \odot Q_2 \colon V_{12}^A \to V_{12}^B$ by
    \begin{align}
        \label{eq:PEPS_Q2Q1_blocks}
        (Q_2 \odot Q_1)_{M_1}^{M_4} &:= \!\! \bigoplus_{M_3 \in \Irr(\mc M^A)} \!\!(Q_2)^{M_4}_{M_3} \otimes (Q_1)_{M_1}^{M_3}
        \, \colon (T_2^A \circ T_1^A)^{M_4}_{M_1} \to (T_2^B \circ T_1^B)^{\sigma'(M_4)}_{\sigma'(M_1)}\, ,
        \\
        \label{eq:PEPS_Q1Q2_blocks}
        (Q_1 \odot Q_2)_{M_1}^{M_4} &:= \!\! \bigoplus_{M_2 \in \Irr(\mc M^A)} \!\! (Q_1)^{M_4}_{M_2} \otimes (Q_2)_{M_1}^{M_2} 
        \, \colon (T_1^A \circ T_2^A)_{M_1}^{M_4} \to (T_1^B \circ T_2^B)_{\sigma(M_1)}^{\sigma(M_4)} \, ,
    \end{align}
    for every $M_1,M_4 \in \Irr(\mc M^A)$, we find by lemma \ref{lemma:PEPS_Q1Q2}, that $Q_{21} = Q_2 \odot Q_1$ and $Q_{12} = Q_1 \odot Q_2$. Together with eq.~\eqref{eq:PEPS_Q21}, lemma \ref{lemma:PEPS_Qv} and the fact that $A^i,B^i$ preserve the labels of the corners of the shape $Y=(1)$, we obtain
    \begin{align*}
        \phi_{21}^A \circ A^i \circ (v_{12}^A)^{M_4}_{M_1}
        &= \phi_{21}^B \circ Q_{21} \circ (A^i \circ v_{12}^A)_{M_1}^{M_4} = \phi_{21}^B \circ (Q_{21} \circ A^i \circ v_{12}^A)^{\sigma'(M_4)}_{\sigma'(M_1)} = \phi_{21}^B \circ B^i \circ (v_{12}^B)_{\sigma'(M_1)}^{\sigma'(M_4)} \, ,
        \\
        \phi_{21}^A \circ A^i \circ (v_{12}^A)^{M_4}_{M_1}
        &= \phi_{21}^B \circ B^i \circ Q_{12} \circ (v_{12}^A)_{M_1}^{M_4} = \phi_{21}^B \circ B^i \circ  (Q_{12} \circ v_{12}^A)^{\sigma(M_4)}_{\sigma(M_1)} = \phi_{21}^B \circ B^i \circ (v_{12}^B)_{\sigma(M_1)}^{\sigma(M_4)} \, .
    \end{align*}
    Cocyclicity of the boundary covectors $(\phB1,\phB2)$ then yields 
    \begin{equation}
        (v_{12}^B)_{\sigma'(M_1)}^{\sigma'(M_4)} = (v_{12}^B)_{\sigma(M_1)}^{\sigma(M_4)} \, . 
    \end{equation}
     For every $M_4$, choose $M_1 \in \Irr(\mc M^A)$ such that the block $(T^A_1 \circ T^A_2)^{M_4}_{M_1}$ is non-zero by non-degeneracy of $\mc M^A$, which implies  $(v_{12}^A)_{M_1}^{M_4} \neq 0$ since it has full support, and $(v_{12}^B)_{\sigma(M_1)}^{\sigma(M_4)} \neq 0$ since $Q_{12}$ is invertible. Bringing everything together, $\sigma = \sigma'$. Graphically, 
    \begin{equation}
        \QSQ{Q1}{}{}{}{} =  
        \! \bigoplus_{M_1,M_2}
        \QSQ{Q1}{M_1}{M_2}{\sigma(M_1)}{\sigma(M_2)}
        \, , \q
        \QSQ{Q2}{}{}{}{} =  
        \! \bigoplus_{M_1,M_2}
        \QSQ{Q2}{M_1}{M_2}{\sigma(M_1)}{\sigma(M_2)} \, ,
    \end{equation}
    which is our lemma.
\end{proof}

\medskip
\noindent
An important consequence of lemma \ref{lemma:PEPS_bijections} is that one can now consider $\odot$-products of the form $Q_1 \odot Q_1$, $Q_2 \odot Q_2$, $Q_1 \odot Q_{12}$, etc. We are now ready to prove the crux of our fundamental theorem.

\begin{proposition} \label{proposition:PEPS_fundamental}
    Under hypothesis \ref{hyp:PEPS}, there are unique invertible maps $Q_{21} \colon V^A_{21} \to V^B_{21}$ and $Q_{12} \colon V^A_{12} \to V^B_{12}$ such that
    \begin{equation}
        \label{eq:PEPS_prop_Q}
        Q_{21} \circ A^i 
        =
        B^i \circ Q_{12}  \, , 
    \end{equation}
    for every $i \in [d]$, and 
    \begin{equation}
        \label{eq:PEPS_prop_Qv}
        Q_{21} \circ v^A_{21} = v^B_{21} \, , \q
        Q_{12} \circ v^A_{12} = v^B_{12} \, , \q
        \phi^B_{21} \circ Q_{21}  = \phi^A_{21} \, , \q
        \phi^B_{12} \circ Q_{12}  = \phi^A_{12} \, .
    \end{equation}
\end{proposition}
\noindent
\begin{proof}
    Gluing the shape $Y' = (3,1,1)$ to the shape $Y = (2,1)$ from the top-left shows that $X_2X_2X_1X_1$ is exposed in both $X_2X_2X_1X_1 = {\rm t}((2,1))$ and $X_1X_2X_2X_1X_1X_2 = {\rm s}((3,1,1))$.
    Proposition \ref{prop:PEPS_Q} thus yields an invertible map $Q_{2211} \equiv Q_{X_2X_2X_1X_1} \colon V_{2211}^A  \xrightarrow{\sim} V^B_{2211}$. Applying eq.~\eqref{eq:PEPS_Q_action} to $Y= (2,1)$, and then twice to $Y=(1)$, yields
    \begin{equation}
    \begin{split}
        &Q_{2211} \circ (\id_{\VA2} \odot A^{i_1} \odot \id_{\VA1}) \circ \big((A^{i_2} \circ v_{12}^A) \odot (A^{i_3} \circ v_{12}^A)\big) 
        \\
        & \q =
        (\id_{\VB2} \odot B^{i_1} \odot \id_{\VB1}) \circ \big((B^{i_2} \circ v_{12}^B) \odot (B^{i_3} \circ v_{12}^B) \big) 
        \\
        & \q =         (\id_{\VB2} \odot B^{i_1} \odot \id_{\VB1}) \circ \big((Q_{21} \circ A^{i_2} \circ v_{12}^A) \odot (Q_{21} \circ A^{i_3} \circ v_{12}^A) \big) \, ,
    \end{split}
    \end{equation}
    for every $i_1,i_2,i_3 \in [d]$.
    By lemma \ref{lemma:PEPS_Q1Q2},
    \begin{equation}
    \begin{split}
        \PEPSQ{1}{i_1}{i_3}{i_2}{Q21}{}{1}
        \!\! = \PEPSQ{1}{i_1}{i_3}{i_2}{QQ}{Q1Q2r}{1}
        \!\! = \PEPSQ{1}{i_1}{i_3}{i_2}{QQ}{Q12}{1} \, ,
    \end{split}
    \end{equation}
    for every $i_1,i_2,i_3 \in [d]$. Invoking the cyclicity of the boundary vectors $(\vA1,\vA2)$, one extracts the identity
    \begin{equation}
        \label{eq:PEPS_Q2211}
        Q_{2211} \circ (\id_{\VA2} \odot A^{i_1} \odot \id_{\VA1}) = Q_2 \odot ( B^{i_1} \circ Q_{12} ) \odot Q_1 \, .
    \end{equation}
    Consider the r.h.s. of eq.~\eqref{eq:PEPS_Q2211}. By lemma \ref{lemma:PEPS_bijections}$, Q_2 \odot ( B^{i_1} \circ Q_{12} ) \odot Q_1$ acts factor by factor on a $\odot$-product of vectors in $V_{2121}^A \cong \VA2 \odot V^A_{12} \odot \VA1$. Therefore,
    lemma \ref{lemma:PEPS_Qv} together with $Q_{12} = Q_1 \odot Q_2$ yields
    \begin{equation} 
    \begin{split}
        \label{eq:PEPS_Q2211v}
        Q_{2211} \circ \big(\id_{\VA2} \odot (A^{i_1} \circ v^A_{12}) \odot  \id_{\VA1} \big) 
        &= Q_2 \odot (B^{i_1} \circ v_{12}^B) \odot Q_1 
        \\
        &= (Q_2 \odot Q_{21} \odot Q_1) \circ \big(\id_{\VA2} \odot (A^{i_1} \circ v^A_{12}) \odot  \id_{\VA1} \big) \, ,
    \end{split}
    \end{equation}
    where $\id_{\VA2} \odot (A^{i_1} \circ v^A_{12}) \odot  \id_{\VA1} \colon \VA2 \otimes \VA1 \to V^A_{2211}$ denotes the map $v' \otimes v \mapsto v' \odot (A^{i_1} \circ v_{12}^A)\odot v$.
    In the last step of eq.~\eqref{eq:PEPS_Q2211v}, we made use of proposition \ref{prop:PEPS_Q}, which stipulates that $B^{i_1} \circ v_{12}^B = Q_{21} \circ A^{i_1} \circ v_{12}^A$, together with lemma \ref{lemma:PEPS_bijections}.\footnote{Recall that $\odot$-products of maps such as $Q_2 \odot Q_{21} \odot Q_1$ are only meaningful because lemma \ref{lemma:PEPS_bijections} holds.}
    Cyclicity of the boundary vectors $(\vA1,\vA2)$ gives $Q_{2211} = Q_2 \odot Q_{21} \odot Q_1$.
    Substituting back into eq.~\eqref{eq:PEPS_Q2211}, invertibility of $Q_1$ and $Q_2$ together with non-degeneracy of $\mc M^A$ yields
    \begin{equation}
        Q_{21} \circ A^i 
        = \PEPSC{1}{i}{}{}{}{Q21}{}{0}
        = \PEPSC{1}{i}{}{}{}{Q1Q2}{}{0}
        = \PEPSCB{1}{i}{}{}{}{Q1Q2r}{}{0} 
        = \PEPSCB{1}{i}{}{}{}{Q12}{}{0}
        = B^i \circ Q_{12} \, ,
    \end{equation}
    for every $i \in [d]$. 
    Finally, lemma \ref{lemma:PEPS_Qv} together with $Q_{12}= Q_1 \odot Q_2$ and $Q_{21}=Q_2 \odot Q_1 $ provides the action of $Q_{21}$ and $Q_{12}$ on the boundary vectors and covectors. Uniqueness follows from the fact that eqs.~\eqref{eq:PEPS_prop_Q}--\eqref{eq:PEPS_prop_Qv} imply $Q_{21} \circ A^i \circ v_{12}^A = B^i \circ v_{12}^B$ and $\phi_{21}^B \circ B^i \circ Q_{12} = \phi^A_{21} \circ A^i$, for every $i \in [d]$.\footnote{Notice that this proposition requires neither hypothesis \ref{hyp:PEPS} in its full generality, nor the full power of proposition \ref{prop:PEPS_Q}. Indeed, in practice we are only relying on shapes contained within shape $(3,3,3)$.}
\end{proof}

\medskip \noindent
Specialising to the case $\mc M^A \simeq \mc M^B \simeq \Vect$, for which $\odot=\otimes$, proposition \ref{proposition:PEPS_fundamental} reduces  to the ordinary fundamental theorem \ref{th:PEPS_Vec_fundamental}. In the general case, however, proposition \ref{proposition:PEPS_fundamental} does not tell the whole story: it leaves implicit the nature of the invertible maps $Q_{21}$ and $Q_{12}$, and only establishes the intertwining property \eqref{eq:PEPS_prop_Q} for the shape $(1)$.
Both the structure of these maps and the intertwining property for arbitrary shapes are  captured by the following concise statement, which makes full use of our category theoretic formulation.  

\begin{theorem}[Fundamental theorem]
    Under hypothesis \ref{hyp:PEPS}, there is an equivalence $(\mc Q,\zeta) \colon \mc M^A \to \mc M^B$ of $\Fsq$-module categories that preserves the boundary data. It is unique up to isomorphism of $\Fsq$-module functors. 
\end{theorem}
\noindent
\begin{proof}
    For every word $\fr X \in \{X_1,X_2\}^*$, let
    \begin{equation}
        Q_\fr X \colon V^A_{\fr X} \to V^B_{\fr X}
    \end{equation}
    be the $\odot$-product of copies of $Q_1$ and $Q_2$, one for each letter of $\fr X$, such that $(Q_\emptyset)_M^M = \id_{\mathbb C}$ for every $M \in \Irr(\mc M^A)$ and $\emptyset$ the empty word. By lemma \ref{lemma:PEPS_bijections}, this coincides with the map of proposition~\ref{prop:PEPS_Q} for $\fr X=X_1,X_2,X_1X_2,X_2X_1$; moreover it is invertible and
    \begin{equation}
        (Q_\fr X)_{M_1}^{M_2} \colon (T^A_\fr X)_{M_1}^{M_2} \to (T^B_\fr X)_{\sigma(M_1)}^{\sigma(M_2)} \, .   
    \end{equation}
    Take $\mc Q \colon \mc M^A \to \mc M^B$ to be the functor that sends every $M$ to $\sigma(M)$, where $\sigma \colon \Irr(\mc M^A) \to \Irr(\mc M^B)$ is the bijection from lemma \ref{lemma:PEPS_bijections}. Besides, set $\zeta \colon \mc Q \circ \rho^A(-) \Rightarrow \rho^B(-) \circ \mc Q$ to be the natural transformation with components 
    \begin{equation}
        \zeta_{\fr X,M_1} := \!\! \bigoplus_{M_2 \in \Irr(\mc M^A)} \!\! (Q_\fr X)_{M_1}^{M_2} \otimes \id_{\sigma(M_2)}     
    \end{equation}
    for every $\fr X \in \{X_1,X_2\}^*$ and $M_1 \in \Irr(\mc M^A)$. Clearly, $\zeta$ is an invertible transformation natural in $M_1$. Moreover, it is compatible with the monoidal structure since
    \begin{align}
    \nn
        \zeta_{\fr X \fr X',M_1} 
        &= 
        \bigoplus_{M_3} \,  
        (Q_{\fr X} \odot Q_{\fr X'})_{M_1}^{M_3} \otimes \id_{\sigma(M_3)}
        = 
        \bigoplus_{M_2,M_3}   
        (Q_{\fr X})^{M_3}_{M_2} \otimes (Q_{\fr X'})_{M_1}^{M_2} \otimes \id_{\sigma(M_3)}
        \\ \nn 
        &= \Big( \bigoplus_{M_2,M_3} \id_{( T^B_{\fr X})_{\sigma(M_2)}^{\sigma(M_3)}} \otimes (Q_{\fr X'})_{M_1}^{M_2} \otimes \id_{\sigma(M_3)}\Big) 
        \circ 
        \Big( \bigoplus_{M_2,M_3}  (Q_\fr X)_{M_2}^{M_3} \otimes \id_{(T^A_{\fr X'})_{M_1}^{M_2}} \otimes \id_{\sigma(M_3)} \Big) 
        \\
        &=T^B_{\fr X}(\zeta_{\fr X',M_1}) \circ \zeta_{\fr X,T^A_{\fr X'}(M_1)} \, ,
    \end{align}
    for every $\fr X, \fr X' \in \{X_1,X_2\}^*$. 
    It remains to show that $\zeta$ is natural in $\fr X$. Every morphism in $\Fsq$ is a $\mathbb C$-linear combination of tensor products of identities and morphisms given by $e_\fr i$ with $\fr i \in [d]^Y$, $Y \in \Yc$, so it suffices to show that, for every shape $Y \in \Yc$,
    \begin{equation}
        Q_{{\rm t}(Y)} \circ A^\fr i = B^\fr i \circ Q_{{\rm s}(Y)} \, ,
    \end{equation}
    for every $\fr i \in [d]^Y$. This follows from writing $e_\fr i$ as a composite of tensor products of identities and generators $e_i$, $i \in [d]$, using that $Q_{\fr X \fr X'} = Q_\fr X \odot Q_{\fr X'}$ and proposition \ref{proposition:PEPS_fundamental}.
    Since $\sigma$ is a bijection, $\mc Q$ is an equivalence, and so is $(\mc Q,\zeta)$ according to definition \ref{def:PEPS_equivalence}. Moreover, $Q_\fr X$ coincides with the map of proposition~\ref{prop:PEPS_Q} for every non-empty word $\fr X$. Finally, this equivalence of $\Fsq$-module categories preserves the boundary data. Indeed, by construction $\mc Q(M^A_\oplus) = M^B_\oplus$ and $Q_\fr X$ can be recovered from $(\mc Q,\zeta)$ as $Q_\fr X = \zeta_{\fr X,M^A_\oplus} \circ \mc Q(-)$. Moreover, by lemma \ref{lemma:PEPS_Qv}, $Q_1 \circ \vA1 = \vB1$, $Q_2 \circ \vA2 = \vB2$, $\phB1 \circ Q_1  = \phA1$ and $\phB2 \circ Q_2  = \phA2$.
\end{proof}

\medskip \noindent
\begin{remark}
    By construction, hypothesis \ref{hyp:PEPS} only involves tensor network states whose underlying graph is a shape, i.e.\ a connected skew Young diagram. This restriction is what forced us to establish $Q_{21} = Q_2 \odot Q_1$ and $Q_{12} = Q_1 \odot Q_2$ before we could prove the intertwining property \eqref{eq:PEPS_prop_Q}. A stronger assumption would require the equality of states associated with arbitrary connected patches of the square lattice. Any factor of the target of a shape would then be exposed, thereby widening the scope of proposition \ref{prop:PEPS_Q}. In particular, the limitation pointed out in footnote \ref{foot:PEPS_X2X1} would be lifted: $X_2X_1$ would now be exposed in $X_2 X_2 X_1 X_1$, so that the intertwining property \eqref{eq:PEPS_prop_Q} would simply follow from gluing a single box into the notch of the shape $\nicefrac{(2,2)}{(1)}$. Note, however, that establishing $Q_{21} = Q_2 \odot Q_1$ and $Q_{12} = Q_1 \odot Q_2$ would still be required in order to prove that the maps $Q_\fr X$ assemble into an equivalence of module categories.
\end{remark}

\noindent
The results in this manuscript could be generalised in various ways. First of all, we expect the same strategy to apply to tensor network representations on the hexagonal geometry or associated with more general monoidal signatures, the main difficulties being the presence of several building-block tensors, and finding a suitable description of the corresponding shapes. The content of the resulting fundamental theorem would be largely unchanged. Another possibility is to weaken hypothesis \ref{hyp:PEPS} in various ways. The first option is to allow for more general boundary data, e.g. encoded into boundary tensor network states with distinct entanglement. In that case, one may expect maps $Q_\mc X$ to assemble into a module functor that is not necessarily an equivalence. A similar direction consists in only requiring the states to be proportional to each other, with constants that cannot be absorbed into a rescaling of the tensors and/or of the boundary data. By analogy with tensor network representations of string-net states \cite{Lootens:2020mso}, one may also expect such a weakening to allow for maps $Q_\mc X$ to yield more general module functors. Finally, one could only require equality of states on a fixed shape, presumably the shape $(3,3,3)$. By analogy with the one-dimensional case, we expect this to require the 2-injectivity condition \ref{def:2injectivity} and to only yield an intertwining property of the form \eqref{eq:PEPS_prop_Q} only up to complex constants, which would spoil the equivalence of the module categories.

\newpage

\bibliographystyle{mod-alpha-url}
\bibliography{refs}  

\end{document}